\documentclass[aps,prb,twocolumn,showpacs,floatfix]{revtex4}
\usepackage{dcolumn}% Align table columns on decimal point
\usepackage{bm}% bold math
\usepackage{amsmath,amssymb,graphicx}

\def\unitangstrom{\,\textrm{\AA}}
\def\unitev{\,{\rm eV}}

\def\acc{a_{\rm cc}}
\def\uA{{\bf u}_{\rm A}}
\def\uB{{\bf u}_{\rm B}}
\def\ur{{\bf u}({\bf r})}
\def\xhat{{\hat{\bf e}_x}}
\def\dgamma0{\delta\gamma_0}
\def\bsigma{{\bf \sigma}}
\def\vfermi{v_{\rm F}}
\def\Smax{S_{\rm max}}
\def\Melph{M_{\rm ep}}

\def\Helph{H_{\rm ep}}
\def\uarm{u_{\rm arm}}
\def\gon{g_{\rm on}}
\def\goff{g_{\rm off}}
\def\thk{\Theta({\bf k})}
\def\imag{{\rm i}}
\def\euler{{\rm e}}

\begin{document}
\preprint{Submitted to Phys. Rev. B}

%%%%%%%%%%%%%%%%%%%%%%%%%%%%%%%%%%%%%%%%%%%%%%%%%%%%%%%%%%%%%%%%%%%%%%%%%%%%%%%
%                          Title and Authors
%%%%%%%%%%%%%%%%%%%%%%%%%%%%%%%%%%%%%%%%%%%%%%%%%%%%%%%%%%%%%%%%%%%%%%%%%%%%%%%

\title{Coherent radial breathing like phonons in graphene nanoribbons}

\author{G.~D.~Sanders$^1$, A.~R.~T.~Nugraha$^2$, R.~Saito$^2$,
C.~J.~Stanton$^1$}

\affiliation{
$^1$Department of Physics, University of Florida, Box 118440,
Gainesville, Florida 32611-8440, USA\\
$^2$Department of Physics, Tohoku University, Sendai
980-8578, Japan}

\date{\today}

%%%%%%%%%%%%%%%%%%%%%%%%%%%%%%%%%%%%%%%%%%%%%%%%%%%%%%%%%%%%%%%%%%%%%%%%%%%%%%%
%                            Abstract
%%%%%%%%%%%%%%%%%%%%%%%%%%%%%%%%%%%%%%%%%%%%%%%%%%%%%%%%%%%%%%%%%%%%%%%%%%%%%%%

\begin{abstract}
We have developed a microscopic theory for the generation and
detection of coherent phonons in armchair and zigzag graphene nanoribbons
using an extended tight-binding model for the electronic
states and a valence force field model for the phonons.  The coherent
phonon amplitudes satisfy a driven oscillator equation with the
driving term depending on photoexcited carrier density.  We examine the
coherent phonon radial breathing like mode
amplitudes as a function of excitation energies and nanoribbon types.
For photoexcitation near the optical absorption edge the coherent
phonon driving term for the radial breathing like mode is much larger
for zigzag nanoribbons where transitions between localized edge states
provide the dominant contribution to the coherent phonon driving term.
Using an effective mass theory, we explain how the armchair nanoribbon
width changes in response to laser excitation.
\end{abstract}

\pacs{63.22.-m, 73.22.-f, 78.67.-n}

% 63.22.-m Phonons or vibrational states in low-dimensional structures
% 73.22.-f Electronic structure of nanoscale materials and related systems
% 78.67.-n Optical properties of low-dimensional structures

%\keywords{}

\maketitle

%%%%%%%%%%%%%%%%%%%%%%%%%%%%%%%%%%%%%%%%%%%%%%%%%%%%%%%%%%%%%%%%%%%%%%%%%%%%%%%
\section{Introduction}
\label{Introduction section}
%%%%%%%%%%%%%%%%%%%%%%%%%%%%%%%%%%%%%%%%%%%%%%%%%%%%%%%%%%%%%%%%%%%%%%%%%%%%%%%

Excited state lattice vibrations in carbon nanotubes have been studied
with coherent phonon (CP) spectroscopy.~\cite{gambetta06-cp,
  lim06-cpexp, lim07-cpkorean, kato08-cpaligned, sanders09-cp} Using
CP spectroscopy with pulse shaping techniques, radial breathing mode
(RBM) coherent phonons in chirality-specific semiconducting
single-walled carbon nanotubes have been studied
experimentally.~\cite{sanders09-cp, kim09-cpprl, booshehri11-cppol,
  lim10-acsnano} The CP signals are resonantly enhanced when the pump
photon energy coincides with an exciton resonance, and provides
information on the chirality-dependence of light absorption, phonon
generation, and phonon-induced band structure modulation.

Recently we developed a microscopic theory for the generation and
detection of coherent phonons in single-walled carbon nanotubes in CP
spectroscopy experiments.~\cite{sanders09-cp, kim09-cpprl,
  booshehri11-cppol} We used the Heisenberg equation to obtain a
driven oscillator equation for the coherent phonon amplitudes and
found that the driving function depends explicitly on the
time-dependent photoexcited carrier distribution
functions.~\cite{sanders09-cp} Comparing theory and experiment we find
that our model predicts overall trends in the relative strength of the
RBM coherent phonon signal both within and between different tube types.

In a followup paper we studied the chirality dependence of coherent
phonon amplitudes in single-wall carbon nanotubes over a large range
of chiralities using an effective mass
approach.~\cite{nugraha11-cpprb} By examining the $k$-dependent
electron-phonon interaction in the effective mass approximation, we
were able to explain why some nanotubes start their coherent RBM
diameter oscillations by initially expanding while others start their
RBM diameter oscillations by initially shrinking their diameters, a
fact recently observed experimentally.~\cite{lim10-acsnano} In many
solids, lattices tend to expand when photoexcited by ultrafast laser
pulses in accordance with the Franck-Condon principle. As pointed out
in Ref.~\onlinecite{nugraha11-cpprb} this is not the case for RBM
diameter oscillations in carbon nanotubes where the diameter can
either expand or contract depending on the nanotube chirality and
photoexcitation energy.

The electronics industry is exploring device technologies based on
carbon nanotubes, graphene, and graphene nanoribbons
(GNRs).~\cite{avouris07-cntrev} Field-effect transistors based on GNRs
have been demonstrated~\cite{chen07-gnr} and it is now possible to
fabricate GNRs with atomically precise widths using a number of
methods.~\cite{fasoli09-gnrwire, tapaszto08-gnrstm, bai09-gnrnlett,
  kosynkin09-gnrcnt, cai10-gnratom, masubuchi09-gnrfabr} An
understanding of the electronic and transport properties of GNRs is
essential in realizing device applications for
GNRs.~\cite{huang09-frontier} In particular it is important for
characterization and transport modeling to have a good understanding
of the electron-phonon interaction and lattice vibrations in GNRs. It
is hoped that coherent phonon spectroscopy will prove useful in
characterizing graphene and graphene nanoribbons in addition to carbon
nanotubes.

In this paper, we extend our coherent phonon theory to the case of
unpassivated zigzag and armchair nanoribbons.  In our
model, we calculate electronic states for the $\pi$ electrons in an
extended tight binding model (ETB),~\cite{sanders09-cp,porezag95}
while the phonon modes are treated in a valence force field
framework.~\cite{saito98-phys, jishi93-phonon}
Details concerning the computation of electronic states and phonon
modes can be found in Appendices \ref{app:electronic states} and
\ref{app:phonon modes} respectively. In our formalism, we
incorporate the electron-phonon interaction,~\cite{jiang05-elphint}
the optical matrix elements,~\cite{gruneis04-phd} and the interaction
of carriers with a classical ultrafast laser
pulse.~\cite{sanders09-cp} For simplicity, we neglect the many-body
Coulomb interaction.  In carbon nanotubes, we found that there were
exactly four CP active phonon modes independent of the tube's
chirality.  The most easily observed CP active mode in nanotubes is
the RBM, which has the lowest frequency. In zigzag and armchair nanoribbons,
on the other hand, the rotational degree of freedom about the nanotube axis
is lost and the number of CP active phonon modes in zigzag or armchair
nanoribbons is equal to the number of AB carbon dimers in the nanoribbon
translational unit cell. The most easily observed of these CP modes is
the one with the lowest frequency, namely the radial breathing like
mode (RBLM).

%%%%%%%%%%%%%%%%%%%%%%%%%%%%%%%%%%%%%%%%%%%%%%%%%%%%%%%%%%%%%%%%%%%%%%%%%%%%%%%
\section{Theory}
\label{Theory section}
%%%%%%%%%%%%%%%%%%%%%%%%%%%%%%%%%%%%%%%%%%%%%%%%%%%%%%%%%%%%%%%%%%%%%%%%%%%%%%%

%
%..............................................................................
% Nanoribbon lattice structure
%..............................................................................
\begin{figure} [tbp]
\includegraphics[width=8cm]{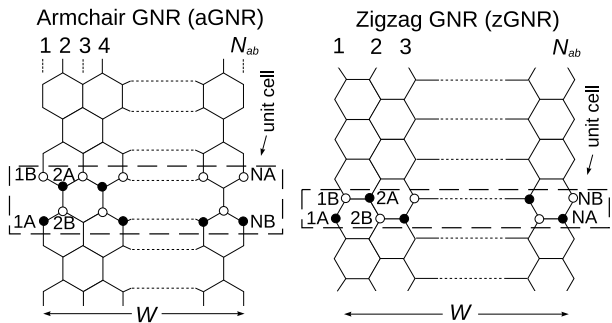}
\caption{Schematic diagram showing lattice structures and translational unit
cells for armchair (aGNR) and zigzag (zGNR) graphene nanoribbons. The width of
the nanoribbons is $W$.}
\label{Nanoribbon structure figure}
\end{figure}
%..............................................................................

We consider two types of nanoribbon, namely zigzag and armchair
ribbons.~\cite{nakada96-gnr, son06-gapgnr} The lattice structure for armchair
and zigzag graphene nanoribbons is shown schematically in
Fig.~\ref{Nanoribbon structure figure}. These ribbons are denoted
$N_{ab} \ \textrm{zGNR}$ and $N_{ab} \ \textrm{aGNR}$ respectively
where $N_{ab}$ is the number of AB carbon dimers in the translational
unit cell.  In zigzag ribbons, the length $L$ of the translational
unit cell is $a$ and the width $W$ of the ribbon is $\left( N_{ab} - 1
\right) \frac{\sqrt{3}}{2} a$ where $a = 2.49\unitangstrom$ is the
hexagonal lattice constant in graphene.  In armchair ribbons, the
translational unit cell length is $\sqrt{3} a$ and the ribbon width is
$\left( N_{ab} - 1 \right) \frac{1}{2} a$. Note that in zigzag and
armchair ribbons with the same number of atoms per unit cell, the area
of the unit cells are equal. Note that we do not allow for the relaxation
of the bond lengths at the surfaces.

%------------------------------------------------------------------------------
\subsection{Coherent phonon generation}
\label{coherent phonon generation subsection}
%------------------------------------------------------------------------------

In coherent phonon spectroscopy, the coherent phonon modes that
are usually excited are the modes with wavevector $q = 0$ whose amplitudes
satisfy a driven oscillator equation
\begin{equation}
\label{Coherent phonon driven oscillator equation}
\frac{\partial^2 Q_{m}(t)}{\partial t^2} +
\omega^2_m Q_{m}(t) = S_m(t) ,
\end{equation}
where $m$ denotes the phonon mode and $\omega_m \equiv \omega_{m}(q=0)$ is the
frequency of phonon mode $m$ at $q = 0$.
There is no damping term in Eq.~(\ref{Coherent phonon driven oscillator equation})
since anharmonic terms in the electron-phonon Hamiltonian are neglected.  We
solve the driven oscillator equation subject to the initial conditions
$Q_m(0) = \dot{Q}_m(0) = 0$.  The driving function $S_m(t)$ is given by
\begin{equation}
\label{Coherent phonon driving function}
S_m(t) = -\frac{2\omega_m}{\hbar} \sum_{nk}
M^m_{n}(k) \left( f_{n}(k,t) - f^0_{n}(k) \right).
\end{equation}
where $f_{n}(k,t)$ are the time-dependent electron distribution
functions and $f^0_{n}(k)$ are the initial equilibrium electron
distribution functions. Here $n$ labels the electronic state
and $k$ is the electron wavevector. The electron-phonon matrix
element $M^m_{n}(k) \equiv M^{m0}_{nk;nk}$ where $M^{mq}_{n'k';nk}$
is the deformation potential electron-phonon matrix element in our
ETB model. Details concerning calculation of this matrix element
can be found in Appendix \ref{app:elph ETB}.

The driving function $S_m(t)$ depends on the photoexcited electron
distribution functions which can be calculated in the Boltzmann
equation formalism taking photogeneration and relaxation effects into
account. In CP spectroscopy, an ultrafast laser pulse generates
electron-hole pairs on a time scale short in comparison with the
coherent phonon period. The observed CP signal is proportional to the
power spectrum of $Q_m(t)$. For simplicity we ignore relaxation effects
and retain only the rapidly varying photogeneration term in the Boltzmann
equation. This model works best when the carrier relaxation time is greater
than the coherent phonon period. The ultrafast pump pulse gives an impulsive
kick to $Q_m(t)$ setting the coherent phonon oscillations in motion.
The subsequent slow relaxation of $S_m(t)$ shifts the coordinate about which
$Q_m(t)$ oscillates but the coherent phonon oscillator is able to
follow this shift adiabatically with negligible change in the oscillation
amplitude and the corresponding power spectrum at the oscillation frequency. Thus
neglecting carrier relaxation has a negligible effect on the computed CP signal.

The photogeneration rate is obtained from Fermi's golden rule and the
equation of motion for the distribution functions on time scales short
in comparison to the relaxation time is
\begin{eqnarray}\label{Photogeneration rate}
\nonumber &&
\frac{\partial f_{n}(k)}{\partial t} =
\frac{8 \pi^2 e^2 \ u(t)}{\hbar \ n_g^2 \ (\hbar\omega)^2}
\left(\frac{\hbar^2}{m_0} \right) \sum_{n'}
\left| P_{n n'}(k) \right|^2
\\ &&
\times \Big( f_{n'}(k,t) - f_{n}(k,t) \Big)
\ \delta \Big( \Delta E_{n n'}(k) - \hbar\omega \Big) ,
\end{eqnarray}
where $\Delta E_{n n'}(k) = \arrowvert E_{n}(k) - E_{n'}(k)
\arrowvert$ are the $k$ dependent transition energies, $\hbar\omega$
is the pump energy, $u(t)$ is the time-dependent energy density of the
pump pulse, $e$ is the electron charge, $m_0$ is the free electron
mass, and $n_g$ is the index of refraction in the surrounding
medium. The pump energy density $u(t)$ is related to the fluence by $F
= \int dt \ u(t) \ (c / n_g)$. The pump energy density is taken to be
a Gaussian with an intensity full width at half maximum (FWHM) of
$\tau_p$ which we define as the pump duration. The optical dipole
matrix element is given by
\begin{equation}\label{Optical matrix element}
P_{n n'}(k)= - \frac{i \hbar}{\sqrt{2 m_0}}\ \hat{\epsilon} \cdot \int
d\textbf{r} \ \psi_{nk}^{*}(\textbf{r}) \ \nabla
\psi_{n'k}^{}(\textbf{r}) ,
\end{equation}
where $\hat{\epsilon}$ is the complex electric polarization vector of
unit length and $\psi_{nk}(\textbf{r})$ are the $\pi$ electron tight-binding
wavefunctions defined in Appendix \ref{app:electronic states}.
With the aid of the $\pi$ orbital expansions of the electron wavefunctions
we can evaluate the optical dipole matrix element in
Eq.~(\ref{Optical matrix element}) analytically. To account for
spectral broadening of the laser pulses we replace the delta function in
Eq.~(\ref{Photogeneration rate}) with a Lorentzian
lineshape~\cite{chuang95}
\begin{equation}\label{Delta function broadening}
\delta(\Delta E-\hbar\omega) \rightarrow
\frac{\Gamma_p /(2\pi)} {{(\Delta E-\hbar\omega)^2+(\Gamma_p/2)^2}} ,
\end{equation}
where $\Gamma_p$ is the FWHM spectral linewidth of the pump pulse.

In addition to the photogeneration rate, there are also carrier relaxation
effects to consider. In Ref.~\onlinecite{dawlaty2008measurement}
Dawlaty \textit{et al.}  measured carrier relaxation times in
photoexcited graphene using pump-probe spectroscopy and found two
relaxation time scales for relaxation of photogenerated carriers.  An
initial fast relaxation transient with a relaxation time ranging from
70 to 120~fs is followed by a slower relaxation with a relaxation time
ranging from 400 to 1700~fs.  These effects can be modeled in a
phenomenological relaxation time approximation which adds a term to
the photogeneration rate in Eq.~(\ref{Photogeneration rate}) of the
form
\begin{equation}\label{Relaxation Rate}
\left( \frac{\partial f_{n}(k)}{\partial t} \right)_{\mbox{relax}}
= - \ \left( \frac{f_{n}(k,t) - f^0_n(k)}{\tau_{r}} \right)
\end{equation}
where $\tau_{r}$ is the phenomenological relaxation time and $f^0_{n}(k)$ are the
initial carrier distribution functions in thermal equilibrium.

In our tight-binding model we only include $\pi$ bands. The $\sigma$
bands in graphene have a direct gap at the $\Gamma$ point of around 6
eV.~\cite{saito98-phys} In nanoribbons the $\sigma$ band direct gap
will be even larger due to quantum confinement effects. The laser
pulses discussed in this work are restricted to photon energies $\hbar
\omega \le 5 \ \mbox{eV}$ so we need not consider coherent phonon
generation due to photoexcited carriers from the $\sigma$ bands.

From the coherent phonon amplitudes, the time-dependent macroscopic
displacements of each carbon atom in the nanoribbon are given by
\begin{equation} \label{Coherent phonon displacement}
\textbf{U}_{sj}^{l}(t) = \frac{\hbar}{\sqrt{2 \rho L_\Omega}} \sum_{m}
\frac{\hat{\textbf{e}}_{sj}^m}{\sqrt{\hbar\omega_m}} \ Q_{m}(t)
\end{equation}
where $\hat{\textbf{e}}_{sj}^m \equiv \hat{\textbf{e}}_{sj}^m(q=0)$
is defined in Appendix \ref{app:phonon modes}.

%------------------------------------------------------------------------------
\subsection{Coherent phonon detection}
\label{Coherent phonon detection subsection}
%------------------------------------------------------------------------------
In coherent phonon spectroscopy a probe pulse is used to measure the
time-varying absorption coefficient of the nanoribbon. The
time-dependent absorption coefficient is given
by~\cite{chuang95,bassani75}
\begin{equation} \label{Absorption coefficient}
\alpha(\hbar\omega,t) =
\frac{\hbar\omega}{n_g \hbar c} \ \varepsilon_2(\hbar\omega,t) ,
\end{equation}
where $\varepsilon_2(\hbar\omega,t)$ is the imaginary part of the
time-dependent dielectric function evaluated at the probe photon
energy $\hbar\omega$.

The imaginary part of the nanoribbon dielectric function is obtained
from Fermi's golden rule
\begin{eqnarray} \label{Imaginary dielectric function}
\nonumber &&
\varepsilon_2(\hbar\omega) =
\frac{8 \pi^2 e^2}{W L_z (\hbar\omega)^2}
\left( \frac{\hbar^2}{m_0} \right) \sum_{n n'}
\int \frac{dk}{\pi}
\ \arrowvert P_{n n'}(k) \arrowvert^2
\times
\\ &&
\Big( f_{n}(k) - f_{n'}(k) \Big)
\ \delta \Big( E_{n'}(k) - E_{n}(k)- \hbar\omega \Big) ,
\end{eqnarray}
where $W$ is the nanoribbon width and $L_z = 3.4\unitangstrom$ is the
nanoribbon thickness taken to be the interlayer distance in graphite.
We replace the delta function in Eq.~(\ref{Imaginary dielectric function})
with a broadened Lorentzian spectral lineshape with a
FWHM of $\Gamma_s$. The distribution function $f_{n}(k)$ and
bandstructure $E_{n}(k)$ are time-dependent.  The time-dependence of
$f_{n}(k)$ comes from the photogeneration of carriers described by the
Boltzmann carrier dynamics and the
time-dependence of $E_{n}(k)$ comes from variations in the
carbon-carbon bond lengths due to the macroscopic coherent phonon
induced atomic displacements in Eq.~(\ref{Coherent phonon displacement}).
This time-dependent deformation of the
nanoribbon bond lengths alters the tight-binding Hamiltonian and overlap
matrix elements in the extended tight-binding
model.~\cite{sanders09-cp} Note that to first order in the lattice
displacements the energies $E_{n}(k)$ vary with time while the
tight-binding wavefunctions and optical matrix elements $P_{n n'}(k)$
do not.

In coherent phonon spectroscopy, excitation of coherent phonons by the
pump modulates the optical properties of the nanoribbons giving rise
to a transient differential transmission signal. In our model we take
the theoretical CP signal to be proportional to the power spectrum of
the transient differential transmission after background
subtraction. We compute the power spectrum using the Lomb periodogram
algorithm described in Ref.~\onlinecite{press92}.

%%%%%%%%%%%%%%%%%%%%%%%%%%%%%%%%%%%%%%%%%%%%%%%%%%%%%%%%%%%%%%%%%%%%%%%%%%%%%%%
\section{Zigzag nanoribbon results}
\label{Zigzag nanoribbon results section}
%%%%%%%%%%%%%%%%%%%%%%%%%%%%%%%%%%%%%%%%%%%%%%%%%%%%%%%%%%%%%%%%%%%%%%%%%%%%%%%

%------------------------------------------------------------------------------
\subsection{Bandstructure and absorption spectra}
%------------------------------------------------------------------------------

In zigzag nanoribbons, the nanoribbon translational unit cell length
is $L = a = 2.49\unitangstrom$ and the nanoribbon width is $W =\left(
N_{ab} - 1 \right) \frac{\sqrt{3}}{2} a$.  We consider in detail a
zigzag nanoribbon with 7 dimers per translational unit cell
(7~zGNR). In this case $N_{ab}=7$ and the width of the ribbon is $W =
12.94\unitangstrom$. The presence of the edges in the nanoribbon
qualitatively alters the electronic and optical properties.
%
%..............................................................................
% band structure and density of states for 7 zGNR nanoribbon
%..............................................................................
\begin{figure} [tbp]
\includegraphics[scale=.75]{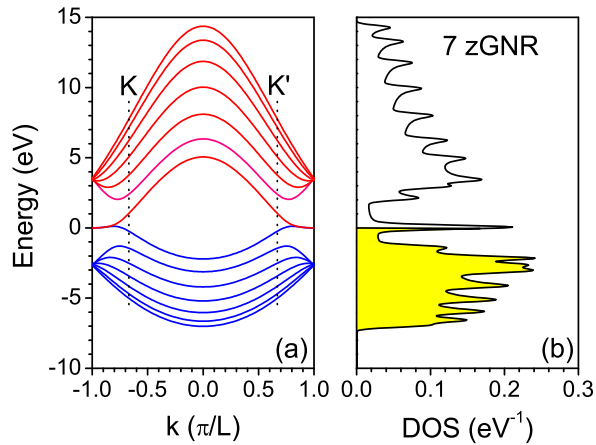}
\caption{ (color online) (a) Computed ETB electronic bands for $\pi$
  electrons in a zigzag nanoribbon with 7 atoms per translational unit
  cell. (b) Corresponding electronic density of states. The density
  of states per carbon atom for the valence bands are shaded
  in yellow. }
\label{Enk 7zGNR}
\end{figure}
%..............................................................................

The computed bandstructure $E_n(k)$ and density of electronic states
$\mbox{DOS(E)}$ for $\pi$ bands in 7~zGNR nanoribbons is shown in
Fig.~\ref{Enk 7zGNR}(a) and (b).  The Brillouin zone is one
dimensional with $|k| \leq \pi/L$. There are total of $2 N_{ab}$ or
fourteen bands. The seven bands with negative energy ($E_n(k) < 0$)
are the valence bands and the seven bands with positive energy are the
conduction bands. From Fig.~\ref{Enk 7zGNR}(a) we note that the
conduction and valence bands are asymmetric about $E=0$. This comes from
the atomic overlap matrix elements in the ETB
formalism which is not present in the simple effective mass model.
We see a strong degeneracy in the bands at $k = \pm
\ \pi/L$ and note that there are two partially degenerate bands near
$E=0$. These bands correspond to localized edge states for $|k|
\gtrsim 2\pi/3L$ where $k = \pm \ 2\pi/3L$ are the Dirac points $K$
and $K'$.  For $|k| \lesssim 2\pi/3L$ the states penetrate into the
interior of the ribbon as $|k| \rightarrow 0$. These edge states are
peculiar to zigzag nanoribbons and have been studied
using effective mass, tight-binding and \textit{ab initio}
theories.~\cite{fujita96-gnr, nakada96-gnr, pisani07-gnr} The
remaining bands are delocalized zone folded quantum confined bands the
lowest of which have parabolic minima near the Dirac points $K$ and $K'$
located at $k=\pm \ 2\pi/3L$.  The density of states shown in
Fig.~\ref{Enk 7zGNR}(b) has a sharp peak near $E=0$ due to the
dispersionless localized edge bands in the vicinity of $k = \pm
\ \pi/L$. As we move away from $E=0$ the density of states contains a
series of von Hove singularities at the extrema of the electronic
bands. As $N_{ab}$ increases, the density of states approaches
$\mbox{DOS(E)}$ in planar graphene.  For large $N_{ab}$ the edge state
contribution to $\mbox{DOS(E)}$ decreases as
$1/N_{ab}$.~\cite{nakada96-gnr}
%
%..............................................................................
% Absorption coefficient and transition diagram for 7 zGNR nanoribbon
%..............................................................................
\begin{figure} [tbp]
\includegraphics[scale=.75]{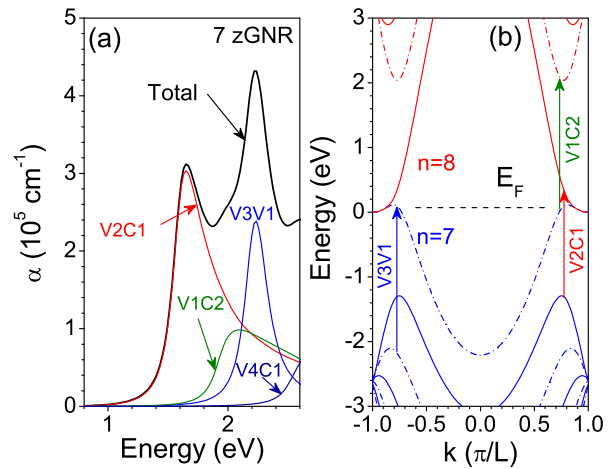}
\caption{ (color online) (a) Band edge absorption for 7~zGNR
  nanoribbons with electric polarization vector parallel to the
  ribbon. Total absorption is the sum of absorption due to several
  transitions. (b) Band diagram showing the Fermi level $E_{\rm F}$ and
  transitions involved in the three lowest absorption peaks.  Blue
  lines are valence bands and red lines are conduction bands. Solid
  lines ($n$ even) are even parity states and the dash-dotted lines
  ($n$ odd) are odd parity states.  }
\label{Optical Transitions 7zGNR}
\end{figure}
%..............................................................................

The absorption spectrum is shown in Fig.~\ref{Optical Transitions 7zGNR}(a)
for the 7~zGNR nanoribbon. For now we assume the light is linearly
polarized with the electric polarization vector parallel to the
nanoribbon's length. Later we will study the in-plane polarization dependence.
The nanoribbon is assumed to be undoped at a
temperature of 300~K.  The Fermi level $E_{\rm F}$ at 70.16~meV is indicated
by the black dashed line in Fig.~\ref{Optical Transitions 7zGNR}(b). The
absorption coefficient is computed using Fermi's
golden rule (Eq.~(\ref{Absorption coefficient})) assuming Lorentzian
lineshapes with a FWHM of $0.2\ \textrm{eV}$. The total absorption
spectrum is shown in Fig.~\ref{Optical Transitions 7zGNR}(a) as a
thick black line. This is a sum of contributions from several
transitions.  The lowest lying absorption peak at 1.65~eV, labeled
V2C1, is due to transitions between the low lying second hole band V2
and the higher lying first electron band C1. This transition is
indicated by a vertical arrow in Fig.~\ref{Optical Transitions 7zGNR}(b)
whose length is the absorption peak transition energy,
\textit{i.e.} 1.65~eV.  From Fig.~\ref{Optical Transitions 7zGNR}(b),
we see that the V2C1 peak comes from transitions between states near
the $K$ and $K'$ points in the Brillouin zone. The initial states are
quantum confined hole states V2 and the final states are the localized
electron edge states C1.  A second broad absorption transition V1C2
peaking at 2~eV comes from transitions between the localized hole edge
states V1 and the second quantum confined electron states C2. There is
a strong peak in the absorption spectrum at 2.23~eV due to the
intraband transition V3V1 between V3 quantum confined states near the
$K$ and $K'$ points and the localized V1 hole edge states.

The absorption spectrum in the 7~zGNR nanoribbon is qualitatively
different from the absorption spectrum seen in nanotubes. A mirror
plane runs down the center of the nanoribbon so the electronic states
are either symmetric or antisymmetric about the mirror plane. The
bands with even parity in Fig.~\ref{Optical Transitions 7zGNR}(b) are
solid lines and the odd parity bands are shown as dash-dotted
lines. The allowed transitions for light polarized parallel to the
nanoribbon axis are between bands with the same parity. In particular
the interband $\Delta n = 0$ selection rule which holds for nanotubes
does not hold for nanoribbons. These conclusions are in agreement with
results obtained within the effective mass formalism. \cite{sasaki11-optgnr}
The edge states play an important role in the band edge transitions and
the absorption spectrum is sensitive to the position of the Fermi level.
In Fig.~\ref{Optical Transitions 7zGNR}(b) we see that the V1 edge states
in the vicinity of the Dirac points lie above the Fermi energy. Since
these states are empty, electrons in the lower bands can be photoexcited
into these V1 states and this accounts for the strong intraband V3V1
absorption peak seen at 2.23~eV.

Optical selection rules and absorption in zigzag nanoribbons have been
studied by other authors using nearest-neighbor tight-binding
approaches.~\cite{hsu07-select,sasaki11-optgnr} They neglect overlap
matrix elements so that the conduction and valence bands are
symmetric about $E=0$. For symmetric valence and conduction bands the
absorption peaks V1C2 and V2C1 are degenerate which disagrees with
what we find taking overlap matrix elements into account. Our edge
state bands V1 and C1 are not completely dispersionless and in undoped
7~zGNR nanoribbons we find $E_{\rm F}$ lies below the V1 band for a range of
$k$ values in the Brillouin zone. This gives rise to a strong V3V1
absorption feature which is not predicted in
Refs.~\onlinecite{sasaki11-optgnr} and~\onlinecite{hsu07-select}.

%------------------------------------------------------------------------------
\subsection{Phonon dispersion relations}
%------------------------------------------------------------------------------
%
%..............................................................................
% VFF phonon dispersion relations for 7 zGNR nanoribbon
%..............................................................................
\begin{figure} [tbp]
\includegraphics[scale=.75]{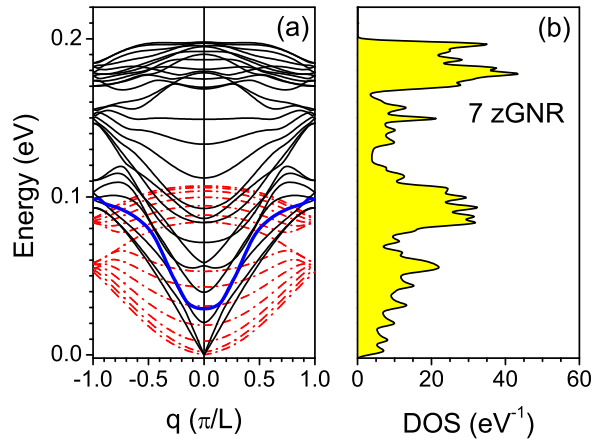}
\caption{ (color online) (a) Phonon dispersion relations for a 7~zGNR
  nanoribbon calculated in the valence force field model. The
  out-of-plane modes are red dash-dotted lines while the remainder are
  in-plane modes. The band associated with the $q=0$ RBLM mode is
  shown as a thick blue line. (b) Density of modes per carbon atom
  for the dispersion relations in (a).  }
\label{Phonon Disperstion 7zGNR}
\end{figure}
%..............................................................................

Phonon dispersion relations for a 7~zGNR nanoribbon obtained in our
valence force field (VFF) model is shown in Fig.~\ref{Phonon
  Disperstion 7zGNR}(a).  The dispersion relations and normal mode
displacements are obtained by solving the dynamical matrix eigenvalue
problem.~\cite{jishi93-phonon} Since the force constant tensor is
block diagonal, the in-plane and out-of-plane modes can be treated
independently.  There are $2\ N_{ab} = 14$ out-of-plane modes and
these are shown as red dash-dotted lines. We find that for normally
incident pump pulses, none of these
out-of-plane modes can be coherently excited and so they are not of
interest to us. The remaining $4\ N_{ab} = 28$ modes are in-plane
modes and are shown as solid lines. We find that $N_{ab}=7$ of the
in-plane $q=0$ modes can be coherently excited. We focus our attention
on the lowest lying of these modes, the so-called radial breathing
like mode (RBLM).  The RBLM mode is a stretching mode in which the
width of the nanoribbon breathes in and out and is somewhat analogous
to the radial breathing mode (RBM) seen in nanotubes. The band
associated with the RBLM is shown in Fig.~\ref{Phonon Disperstion
  7zGNR}(a) as a thick blue line. The phonon energy of the RBLM
coherent phonons is $\hbar\omega = 29\ \mbox{meV}$ which corresponds
to an oscillation period of 0.141~ps.  The density of phonon modes is
shown in Fig.~\ref{Phonon Disperstion 7zGNR}(b).  Apart from numerous
von-Hove singularities, the density of phonon states in
Fig.~\ref{Phonon Disperstion 7zGNR}(b) resembles the density of phonon
states in bulk graphene.~\cite{jishi93-phonon}

The RBLM phonon energies in zGNR nanoribbons obtained from our VFF
model can be fit to $\hbar\omega = A W^P + B$ where $W$ is the zigzag
nanoribbon width in Angstroms and $\hbar\omega$ is the $q=0$ RBLM
phonon energy in eV.  The fitting parameters $A = 0.25247$, $P =
-0.82773$ and $B = -0.00136$ are obtained from a least squares fit to
our VFF results for $3 \leq N_{ab} \leq 15$ corresponding to widths in
the range $4\unitangstrom < W < 30 \unitangstrom$.

%------------------------------------------------------------------------------
\subsection{Generation of coherent phonons}
%------------------------------------------------------------------------------

\begin{figure} [tbp]
\includegraphics[scale=1.]{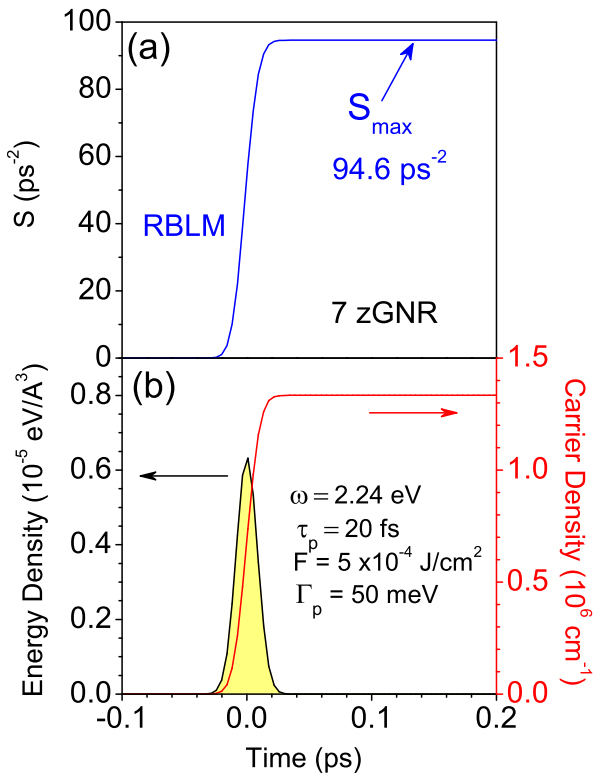}
\caption{ (color online) For 7~zGNR nanoribbon excited by Gaussian
  laser pulse with polarization vector parallel to ribbon length we
  plot (a) coherent RBLM phonon driving function $S(t)$ and (b) pump
  photon energy density (left axis) and photogenerated carrier density
  (right axis). For the assumed pump pulse, the photon energy
  $\omega$, fluence $F$, pulse duration $\tau_p$, and FWHM spectral
  linewidth $\Gamma_p$ are indicated in (b).  }
\label{Smax Excite 7zGNR}
\end{figure}

In a typical simulation, we excite coherent RBLM phonons with a 20~fs
Gaussian laser pulse. In our simulations the fluence F is taken to
be $5 \times 10^4 \ \mbox{J/cm}^2$ and the pump pulse is
linearly polarized with the electric polarization vector parallel to
the nanoribbon length. The pump photon energy is 2.24~eV which
coincides to the V3V1 absorption peak shown in Fig.~\ref{Optical
  Transitions 7zGNR}(a). The pump energy density $u(t)$ is plotted
against the left axis of Fig.~\ref{Smax Excite 7zGNR}(b) and the
photoexcited carrier density per unit length is plotted against the
right axis. The time dependent distribution functions and
photogenerated carrier density are obtained by integrating the
photogeneration rates in Eq.~(\ref{Photogeneration rate}) using the
equilibrium distribution functions as initial conditions. In
integrating this equation, we assume a FWHM pump spectral width of
$\Gamma_p = 150 \ \mbox{meV}$ which appears in the broadened delta function.
%
%..............................................................................
% Photogeneration for 7 zGNR nanoribbon
%..............................................................................
%..............................................................................
%
Using the time-dependent distribution functions, the RBLM coherent
phonon driving function $S(t)$ can be obtained from
Eq.~(\ref{Coherent phonon driving function}). The driving function in our
example is shown in Fig.~\ref{Smax Excite 7zGNR}(a).  It initially rises
sharply during the pump phase and reaches a peak value $\Smax = 94.6
\ \mbox{ps}^{-2}$. Given $S(t)$ we can solve Eq.~(\ref{Coherent phonon
  driven oscillator equation}) to obtain the RBLM coherent phonon
amplitude $Q(t)$. Positive $Q(t)$ for the RBLM mode corresponds to an
expansion of the nanoribbon width while negative $Q(t)$ corresponds to
a contraction. For coherent RBLM phonon oscillations the nanoribbon width
initially expands or contracts depending on whether $\Smax$ is positive
or negative.

%
%..............................................................................
% Relaxation for 7 zGNR nanoribbon
%..............................................................................
\begin{figure} [tbp]
\includegraphics[scale=.9]{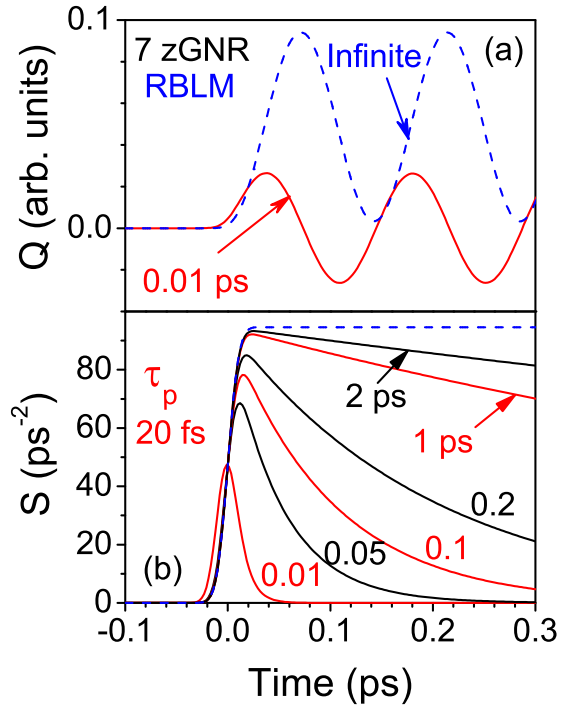}
\caption{ (color online) For 7~zGNR nanoribbon excited by  a 20~fs Gaussian
  laser pulse with polarization vector parallel to ribbon length we plot
  (b) coherent RBLM phonon driving function $S(t)$ for several relaxation times.
  Solid curves are labeled with the relaxation time measured
  in picoseconds. The dashed blue curve is the relaxationless limit.
  In (a) the dashed blue curve shows the coherent phonon amplitude $Q(t)$
  in the relaxationless limit and the solid red curve shows $Q(t)$ assuming
  a 10~fs carrier relaxation time.}
\label{7zGNR S(t) Relaxation}
\end{figure}
%..............................................................................
%
Carrier scattering effects can also effect the strength of the coherent phonon
oscillations especially if the scattering times are short compared to the oscillation
period.  A detailed treatment of carrier scattering is beyond the scope of this article.  However,
we note that carrier scattering by itself does not necessarily lead to a change in the driving
function $S_m(t)$. According to Eq.~\ref{Coherent phonon driving function}, scattering affects
$S_m(t)$\textit{ only if} the carrier scatters to a state that has a \textit{substantially  different} carrier-phonon
matrix element.  For instance, in GaAs, an electron scattering \textit{within} the conduction band
(either via incoherent phonons or electron-electron scattering) does not change the driving function since
the electron-phonon matrix elements are similar within the conduction band.  We would expect that in
graphene nanoribbons, scattering \textit{within} a given band would have less of an effect than scattering
\textit{between} different bands.

We can estimate the effects of carrier dynamics on these results using a simple
relaxation time approximation model for recombination of photogenerated
carriers. We considered relaxation times ranging from
10~fs to 2~ps and our results are shown in Fig.~\ref{7zGNR S(t) Relaxation}.
After an initial sharp rise due to photoexcitation by the pump, $S(t)$ decays
exponentially at long times due to carrier relaxation as seen in
Fig.~\ref{7zGNR S(t) Relaxation}(b). We then solved the driven oscillator
equation (Eq.~\ref{Coherent phonon driven oscillator equation}) for the coherent
phonon amplitude $Q(t)$. In Fig.~\ref{7zGNR S(t) Relaxation}(a) we plot $Q(t)$
assuming a relaxation times of 10~fs (solid red curve) and in the relaxationless
limit (dashed blue curve).

We note that there is a shift in phase of about  $\pi/2$ between the two cases.
For the 10 fs carrier relaxation time, we get a sinusoidal oscillation, while for the no relaxation
case, we find that $Q(t) \propto 1-cos(\omega t)$.  These are the limiting cases
for a driving function that is proportional to a delta function and a step function respectively.  This
is discussed in more detail in Ref.~ \onlinecite{Kuznetsov2001}.

%
%..............................................................................
% Q squared for 7 zGNR nanoribbon
%..............................................................................
\begin{figure} [tbp]
\includegraphics[scale=.7]{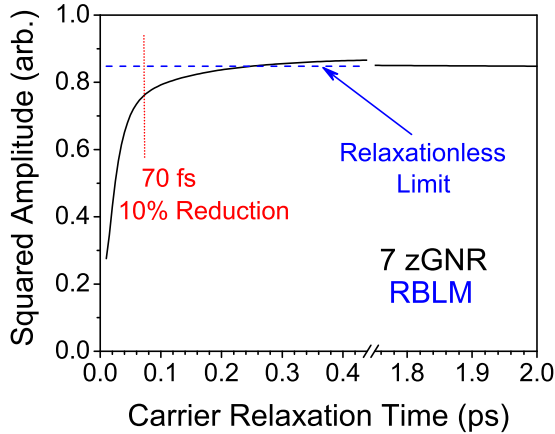}
\caption{ (color online) For 7~zGNR nanoribbon excited by Gaussian
  laser pulse with polarization vector parallel to ribbon length we plot
  the squared RBLM coherent phonon amplitude $Q(t)^2$ as a function of carrier
  relaxation time. The relaxationless limit is shown as a blue dashed line
  for comparison. For a relaxation time of 70~fs the squared amplitude
  $Q(t)^2$ is reduced by 10\%. }
\label{7zGNR Q(t) Squared Relaxation}
\end{figure}
%..............................................................................
%
The CP signal is proportional to the square of the amplitude of the $Q(t)$
oscillations, so to estimate the effect carrier relaxation has on the
computed RBLM CP signal, we solved the driven oscillator equation and found
the amplitude of $Q(t)$ at long times as a function of the carrier relaxation
time. The square of $Q(t)$ is shown in
Fig.~\ref{7zGNR Q(t) Squared Relaxation}. We find that for
relaxation times greater than 200~fs, we recover the results obtained in
the absence of carrier relaxation. This is not surprising since the
RBLM period is 141~fs and the point about which $Q(t)$ oscillates is able to
follow slow changes in $S(t)$ adiabatically while the amplitude of the $Q(t)$
oscillations at long times remains unchanged. For relaxation times less than
the RBLM period, the amplitude of $Q(t)$ at long times decreases, but even
for relaxation times slightly less than the RBLM period the errors are not
too great. For a relaxation time of 70~fs, we see that the square of the $Q(t)$
oscillation amplitude and hence the CP intensity is only reduced by $10\%$.
Since 70~fs is the shortest photoexcited carrier relaxation time measured
in the pump-probe experiments by Dawlaty \textit{et al.} \cite{dawlaty2008measurement}
on graphene samples, we feel that our neglect of carrier relaxation effects
is a fair approximation. Henceforth, we will neglect the effects of carrier relaxation for
simplicity.

\begin{figure} [tbp]
\includegraphics[scale=.8]{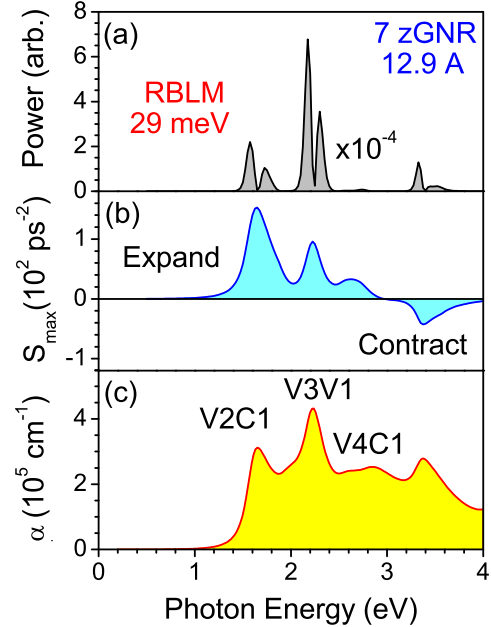}
\caption{ (color online) For 7~zGNR nanoribbon excited by Gaussian a
  laser pulse with polarization vector parallel to ribbon length, we
  plot (a) the coherent phonon power at the RBLM frequency (29~meV),
  (b) the value of $\Smax$, and (c) the initial absorption spectrum as
  a function of photon energy.  }
\label{CP Power 7zGNR}
\end{figure}

Since coherent phonon spectroscopy gives direct phase information on
the coherent phonon amplitude, it is instructive to examine $\Smax$ as
a function of pump photon energy. This is done in Fig.~\ref{CP Power 7zGNR}.
Fig.~\ref{CP Power 7zGNR}(a) is the power spectrum of $Q(t)$ at the RBLM frequency.
In Fig.~\ref{CP Power 7zGNR}(b) we plot $\Smax$ as a function of pump
photon energy. For comparison, the absorption coefficient is plotted in
Fig.~\ref{CP Power 7zGNR}(c). Near the band edge, we see from
Fig.~\ref{CP Power 7zGNR}(b) that the pump light is strongly absorbed
at the V2C1 and V3V1 peaks. The resulting increase in the photoexcited
carrier density increases the coherent phonon driving function and
enhances the coherent phonon oscillation amplitudes. In other words
the coherent phonon driving function near the band edge is determined
by the strength of optical absorption between the lowest few hole
bands and the localized edge states V1 and C1. At energies above 3~eV,
$\Smax$ changes sign and the nanoribbon initially contracts.
%
%..............................................................................
% CP power vs pump energy for 7 zGNR nanoribbon
%..............................................................................
%..............................................................................
%

%------------------------------------------------------------------------------
\subsection{Detection of coherent phonons}
%------------------------------------------------------------------------------

The generation of coherent phonons in zGNR nanoribbons results in
macroscopic oscillations of the carbon atoms. The time-dependent
macroscopic displacement of each carbon atom can be described as a sum
of terms each of which is proportional to a different coherent phonon
amplitude.~\cite{sanders09-cp} These displacements in turn modulate
the optical properties of the nanoribbon through changes induced in
the electronic band structure.
%
%..............................................................................
% Differential gain for 7 zGNR nanoribbon
%..............................................................................
\begin{figure} [tbp]
\includegraphics[scale=.75]{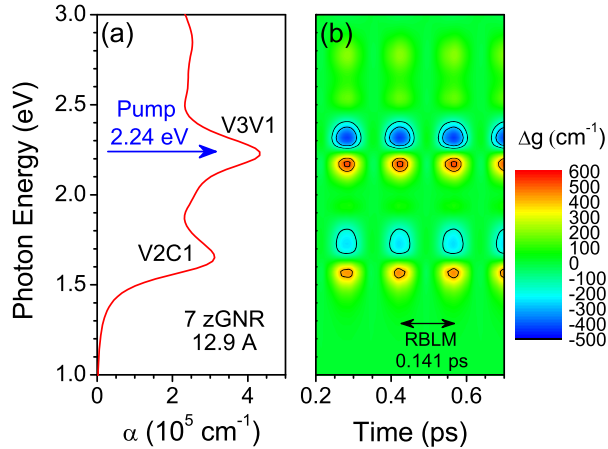}
\caption{ (color online) For 7~zGNR nanoribbon excited by Gaussian
laser pulse with polarization vector parallel to ribbon length, we
plot (a) absorption coefficient as function of photon energy and (b)
differential gain as a function of time and photon energy after the
transient pump pulse dies out. The pump photon energy is taken to be
2.24~eV }
\label{Differential gain 7zGNR}
\end{figure}
%..............................................................................
%

We illustrate our results in Fig.~\ref{Differential gain 7zGNR} where
we simulate the results of a hypothetical two-color pump probe
experiment on a 7~zGNR nanoribbon. As usual the fluence is $5 \times
10^4 \ \mbox{J/cm}^2$ and we assume the pump and probe are both
linearly polarized with electric polarization vector parallel to the
nanoribbon length. As indicated in Fig.~\ref{Differential gain 7zGNR}(a) the
pump photon energy is 2.24~eV which coincides with the peak in the
V3V1 absorption feature. The time-resolved differential gain measured
by the probe is given by
\begin{equation}\label{differential gain}
\Delta g(\hbar\omega,t) =
- \left( \alpha(\hbar\omega,t) - \alpha(\hbar\omega,t\rightarrow-\infty) \right)
\end{equation}
where $\alpha(\hbar\omega,t)$ is the time dependent absorption
coefficient defined in Eq.~(\ref{Absorption coefficient}).
The differential gain is shown in Fig.~\ref{Differential gain 7zGNR}(b) as a
function of probe delay and probe photon energy. There are large
differential gain oscillations near the V2C1 and V3V1 absorption peaks
which oscillate at the RBLM period of 0.141~ps, so it is clear that
the dominant contribution to the time-dependent differential gain
comes from RBLM induced absorption modulation.  We find that the
ribbon width initially expands and begins oscillating about a new and wider
equilibrium width $W_0 + \Delta W$ where $W_0$ is the initial width
before photoexcitation. When the laser pulse duration is
much less than the coherent phonon period, the ribbon width oscillates
between $W_0$ and $W_0 + 2 \ \Delta W$. When the ribbon width $W(t) = W_0$
the differential gain vanishes. When $W(t) = W_0 + 2 \ \Delta W$ the
absorption peaks involving transitions to the localized edge states
shift to higher energies giving rise to positive differential gain on
the low energy side of the absorption peak and negative differential
gain on the high energy side. This is clearly seen in
Fig.~\ref{Differential gain 7zGNR}(b).

The experiments we model are degenerate pump/probe measurements
(i.e. the pump and probe have the same wavelength) and the pump and
probe excitation energy is scanned. After background
subtraction, the time-dependent differential transmission is Fourier
transformed to obtain the power spectrum. Peaks in this power spectrum
occur at the coherent phonon frequencies. Plotting the power spectrum
at the RBLM frequency as a a function of pump/probe photon energy
gives us the RBLM CP power spectrum. The RBLM CP power spectrum is
shown in Fig.~\ref{CP Power 7zGNR}(a) as a function of pump/probe
photon energy. We see strong double peaks in the RBLM CP spectra
corresponding to the V2C1 and V3V1 transitions. Integrating the area
under these double peaks, we obtain the RBLM CP intensity for each
transition.
%
%..............................................................................
% CP power vs polarization angle for 7 zGNR nanoribbon
%..............................................................................
\begin{figure} [tbp]
\includegraphics[scale=.75]{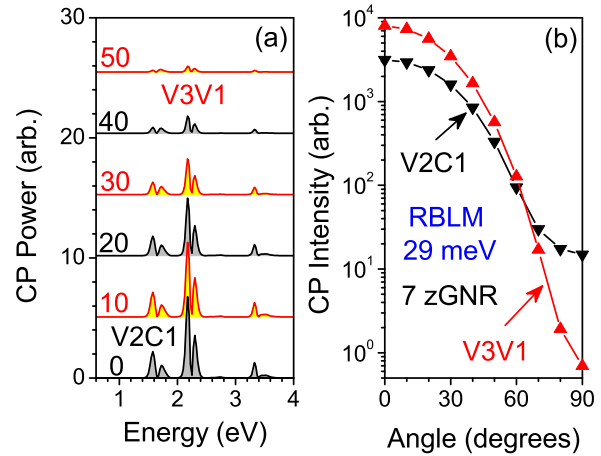}
\caption{ (color online) For 7~zGNR excited by Gaussian laser pulse
  with the pump/probe polarization vector rotated an angle from the
  nanoribbon axis, we plot (a) the RBLM CP spectrum as a function of
  pump/probe energy for several polarization angles from 0 the 50
  degrees and (b) the integrated RBLM CP intensity for V2C1 and V3V1
  transitions.  }
\label{CP polarization 7zGNR}
\end{figure}
%..............................................................................
%

We also investigated the in-plane polarization dependence of the coherent
phonon spectra. Here the polarization of the electric field of both
the pump and probe pulse are rotated by an angle $\theta$ with respect to the
nanoribbon length.
The dependence of the RBLM CP power spectrum on pump/probe
polarization angle is shown in Fig.~\ref{CP polarization 7zGNR}(a)
where the CP power spectra are shown for polarization angles between 0
and 50 degrees. At a polarization angle of 0 degrees, there is a
strong V3V1 signal at 2.25~eV and a weaker V2C1 signal at 1.65~eV. As
the polarization angle increases from 0 degrees to 50 degrees, the CP
signal becomes weaker. The CP intensity is plotted on a log scale in
Fig.~\ref{CP polarization 7zGNR}(b) for the V2C1 and V3V1 transitions.
As the polarization angle varies from 0 to 90 degrees, the CP
intensity is strongly quenched.
%
%..............................................................................
% CP power vs width for zGNR nanoribbons
%..............................................................................
\begin{figure} [tbp]
\includegraphics[scale=1.]{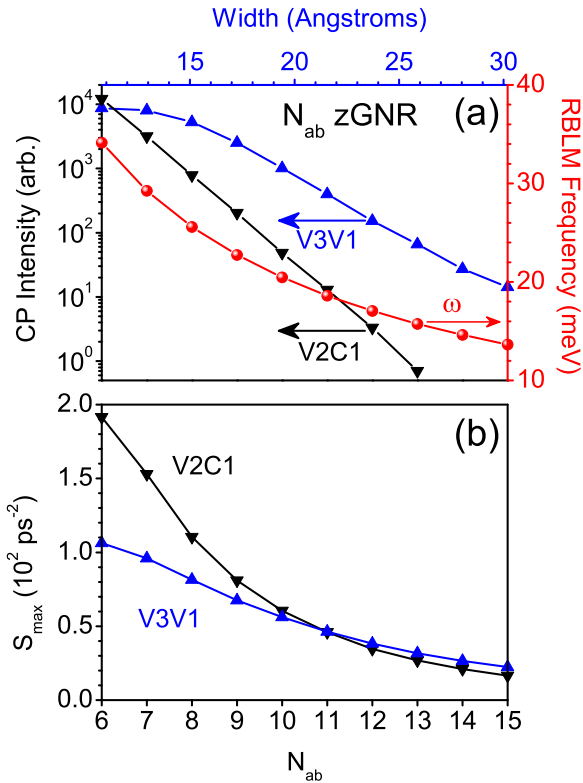}
\caption{ (color online) For zigzag nanoribbons excited by Gaussian
  laser pulse with pump and probe polarization vector parallel to ribbon length,
  in (a) we plot the RBLM CP intensity for the V2C1 and V3V1 transitions as a
  function of $N_{ab}$, the number of carbon dimers in the zigzag unit
  cell, on the left axis. On the right axis in we plot the RBLM frequency
  $\omega$ in eV. In (b) we plot $S_{max}$ for the V2C1 and V3V1 transitions.
  The ribbon width for each value of $N_{ab}$ can be read from the upper axis.  }
\label{zigzag CP intensity vs width}
\end{figure}
%..............................................................................
%

It is useful to plot the CP intensity as a function of nanoribbon
width.  We fix the pump and probe electric polarization vectors to be
parallel to the nanoribbon width as this is the polarization for which
the CP intensity is greatest. In Fig.~\ref{zigzag CP intensity vs
  width}(a), the RBLM CP intensity for the V2C1 and V3V1 transitions
as a function of the number of carbon dimers in the zigzag nanoribbon
unit cell is plotted against the left axis and the RBLM frequency is
plotted against the right axis.  In Fig.~\ref{zigzag CP intensity vs
  width}(b), we plot the coherent driving function amplitude $\Smax$
as a function of $N_{ab}$ for the V2C1 and C3V1 transitions.  We have
studied all zGNR nanoribbons for $N_{ab}$ ranging from 6 to 17 and
find similar results. The driving function is positive for low
energies (up to just below 3~eV) and then becomes negative for all
cases.

%%%%%%%%%%%%%%%%%%%%%%%%%%%%%%%%%%%%%%%%%%%%%%%%%%%%%%%%%%%%%%%%%%%%%%%%%%%%%%%
\section{Armchair nanoribbon results}
\label{Armchair nanoribbon results section}
%%%%%%%%%%%%%%%%%%%%%%%%%%%%%%%%%%%%%%%%%%%%%%%%%%%%%%%%%%%%%%%%%%%%%%%%%%%%%%%

%------------------------------------------------------------------------------
\subsection{Bandstructure and absorption spectra}
%------------------------------------------------------------------------------

In armchair nanoribbons (aGNR), the nanoribbon translational unit cell
length is $L = \sqrt{3} \ a = 4.31\unitangstrom$ and the nanoribbon
width is $W =\left( N_{ab} - 1 \right) \frac{a}{2}$. Armchair
nanoribbons belong to one of three families depending on the mod
number $p = \mbox{mod}(N_{ab},3)$.~\cite{son06-gapgnr,raza08-arm}
Following the convention of Refs.~\onlinecite{son06-gapgnr} and
\onlinecite{raza08-arm} we label the armchair nanoribbons as
$\alpha$-aGNR, $\beta$-aGNR, and $\gamma$-aGNR families
for mod 2, mod 0, and mod 1 nanoribbons respectively.
%
%..............................................................................
% Bandstructures for aGNR nanoribbon families
%..............................................................................
\begin{figure} [tbp]
\includegraphics[scale=.75]{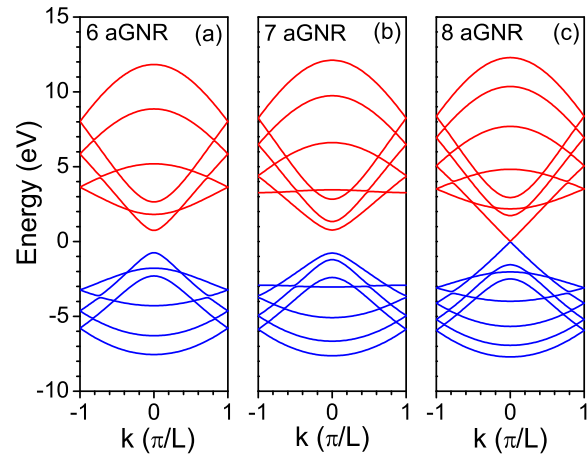}
\caption{ (color online) Electronic band structure for three armchair
  nanoribbons obtained from the ETB model. We have one representative
  from each mod($N_{ab}$,3) armchair nanoribbon family: (a)
  6~$\beta$-aGNR mod 0 semiconducting nanoribbon, (b) 7~$\gamma$-aGNR
  mod 1 semiconducting nanoribbon, and (c) 8~$\alpha$-aGNR mod 2
  metallic nanoribbon.  }
\label{Armchair 3 panel bandstructure}
\end{figure}
%..............................................................................
%

Example bandstructure calculations for each nanoribbon family are
shown in Fig.~\ref{Armchair 3 panel bandstructure}. In our ETB model,
all armchair nanoribbons have a direct gap at $k=0$. From the
6~$\beta$-aGNR mod 0 and 7~$\gamma$-aGNR mod 1 nanoribbon bands in
Fig.~\ref{Armchair 3 panel bandstructure}(a) and (b) we see that
$\beta$- and $\gamma$-aGNR ribbons are semiconducting with parabolic
bands near the band edge. From the 7~$\alpha$-aGNR nanoribbon band in
Fig.~\ref{Armchair 3 panel bandstructure}(c) we see that the
$\alpha$-aGNR ribbons have negligible band gaps and a linear band edge
dispersion relation.  Unlike the case in zigzag nanoribbons, there are
no localized edge states near the band edge. Armchair nanoribbons have
direct gaps that arise from quantum confinement and edge effects and
all the electronic wavefunctions near the band edge are distributed
throughout the width of the ribbon.
%
%..............................................................................
% Band gaps for aGNR nanoribbons
%..............................................................................
\begin{figure} [tbp]
\includegraphics[scale=.75]{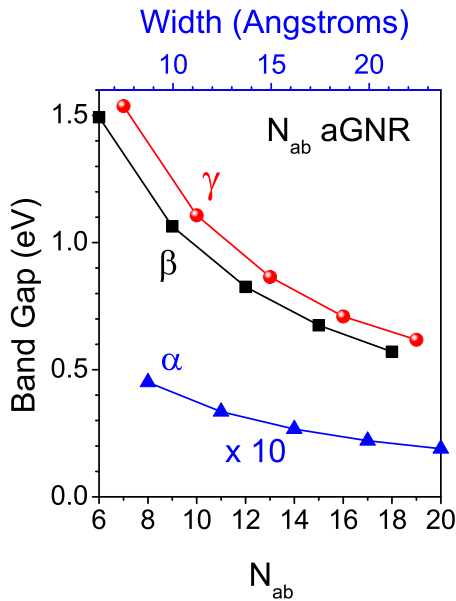}
\caption{ (color online) Direct band gaps for armchair nanoribbons in
the ETB model as a function of the number of dimers $N_{ab}$ in the
translational unit cell. Band gaps are shown for $\alpha$-aGNR mod 2,
$\beta$-aGNR mod 0, and $\gamma$-aGNR mod 1 families.  The band gaps
for the $\alpha$-aGNR mod 2 nanoribbons have been multiplied by a
factor of 10 for clarity. The nanoribbon width is plotted on the top
axis.  }
\label{Armchair band gaps}
\end{figure}
%..............................................................................
%

In Fig.~\ref{Armchair band gaps} we plot the direct band gap $E_g$ in
the ETB model as a function of $N_{ab}$, or equivalently the ribbon
width $W$, for the $\alpha$, $\beta$, and $\gamma$ armchair
nanoribbon families.  In the case of the $\alpha$-GNR family, the
band gap is multiplied by a factor of 10 for clarity. We find that
$\alpha$-aGNRs have a small band gap that decreases with ribbon
width. The $\beta$- and $\gamma$-aGNRs have large band gaps that
decrease as a function of ribbon width, but they fall on different
curves as seen in the figure. Our ETB results are in decent agreement
with first principles results reported in
Ref.~\onlinecite{son06-gapgnr}.  These results differ from simple
tight binding (STB) theory which predicts that band gaps in
$\alpha$-aGNRs vanish identically and that band gaps for $\beta$- and
$\gamma$-aGNRs as a function of ribbon width fall on the same
curve.~\cite{son06-gapgnr}. For each family, we fit the band gaps to
an expression of the form $E_g = A /(W + B)$ where $W$ is the ribbon
width in Angstroms and $E_g$ is the band gap in eV. For wide ribbons
$E_g \approx A/W$.  For the range $6 \unitangstrom < W < 25
\unitangstrom$, fitting parameters for each family are obtained.  For
$\alpha$-aGNRs A = 0.486 eV-$\unitangstrom$ and B = 3.014
$\unitangstrom$, for $\beta$-aGNRs A = 13.8 eV-$\unitangstrom$ and B
= 2.082 $\unitangstrom$, and for $\gamma$-aGNRs A = 14.97
eV-$\unitangstrom$ and B = 2.286 $\unitangstrom$.  These results are
similar to those obtained in Ref.~\onlinecite{raza08-arm} using a
semiempirical extended H\"{u}ckel theory. The armchair nanoribbon
band gaps in Ref.~\onlinecite{raza08-arm} are described by $E_g =
A/W$ where values of A are 0.4, 8.6, and 10.4 eV-$\unitangstrom$ for
$\alpha$- $\beta$- and $\gamma$-aGNRs respectively.

Since there are three distinct armchair nanoribbon families $\alpha$,
$\beta$ and $\gamma$, we've computed room temperature ($T = 300
\ \mbox{K}$) absorption spectra for an undoped nanoribbon in each
family as a function of photon energy for light with electric
polarization vectors parallel to the ribbon length.  In computing the
absorption coefficient using Fermi's golden rule (Eq.~(\ref{Absorption
  coefficient})) we assume Lorentzian lineshapes with the same
$0.2\ \textrm{eV}$ FWHM used in Fig.~\ref{Optical Transitions 7zGNR}.
%
%..............................................................................
% Absorption coefficient and transition diagram for aGNR nanoribbon
%..............................................................................
\begin{figure*} [tbp]
\includegraphics[width=18cm]{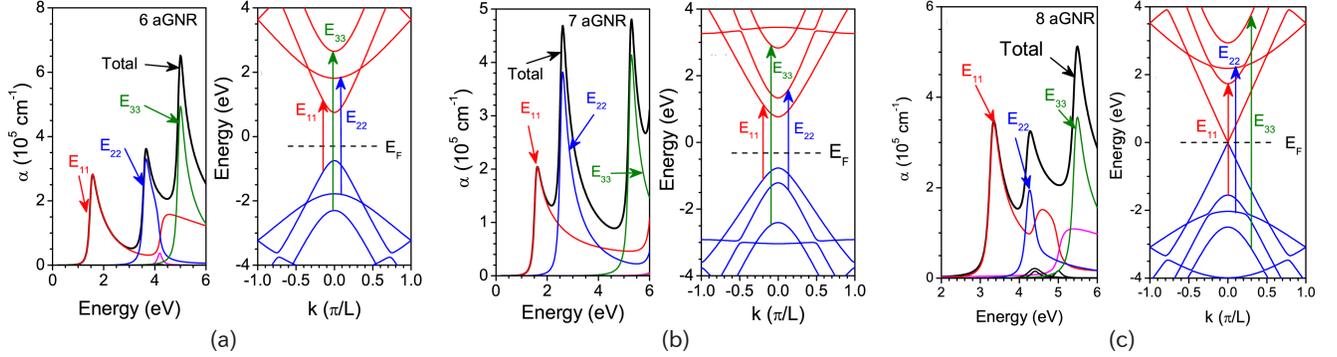}
\caption{ (color online) Band absorptions with electric polarization
  vector parallel to the ribbon and band diagram showing the room
  temperature Fermi level $E_{\rm F}$ with transitions involved in the
  three lowest absorption peaks for (a) 6~$\beta$-aGNR mod 0
  nanoribbon, (b) 7~$\gamma$-aGNR mod 1 nanoribbon, and (c)
  8~$\alpha$-aGNR mod 2 nanoribbon.  Total absorption is the sum of
  absorption due to several transitions.  In the band diagrams, blue
  lines are valence bands and red lines are conduction bands.  }
\label{Optical Transitions aGNR}
\end{figure*}
%..............................................................................

In Fig.~\ref{Optical Transitions aGNR}(a) we plot the absorption
spectrum in an undoped 6~$\beta$-aGNR mod 0 semiconducting nanoribbon
for light polarized parallel to the nanoribbon length. The total
absorption spectrum, shown as a solid black line, is a sum of
absorption spectra due to three major interband direct transitions
labeled $E_{11}$, $E_{22}$ and $E_{33}$. These transitions are shown
in the band diagram in the right panel of Fig.~\ref{Optical
  Transitions aGNR}(a). The sharp peaks in the spectra are due to von
Hove singularities at each of the transition band edges.  The room
temperature Fermi level is $E_{\rm F} = -29.5 \ \mbox{meV}$. The Fermi
level is shifted slightly to the valence band edge since the
conduction and valence bands are asymmetric in the ETB model.  Sasaki
\textit{et al.}~\cite{sasaki11-optgnr} have studied optical selection
rules in armchair nanoribbons and find that for light polarized
parallel to the nanoribbon length, the allowed interband transitions
are direct transitions between valence band $i$ and conduction band
$i$. In order to conform to a widely used convention in the
literature, the transitions in Fig.~\ref{Optical Transitions aGNR}(a)
are labeled $E_{ii}$ as opposed to the V$i$C$i$ notation used in
Section~\ref{Zigzag nanoribbon results section}.  We see that the
optical selection rules in armchair nanoribbons are qualitatively
different from those in zigzag nanoribbons where direct transitions
are \textit{forbidden} for polarization parallel to the
ribbon.~\cite{hsu07-select,sasaki11-optgnr}

In Fig.~\ref{Optical Transitions aGNR}(b) we show similar results for an
undoped 7~$\gamma$-aGNR mod 1 semiconducting nanoribbon. Again the
total absorption spectrum is the sum of absorption spectra due to
three interband direct transitions $E_{11}$, $E_{22}$ and $E_{33}$ and
the room temperature Fermi level $E_{\rm F} = -32.7 \ \mbox{meV}$ is shifted
slightly towards the valence band edge.

In Fig.~\ref{Optical Transitions aGNR}(c) we show results for an
undoped 8~$\alpha$-aGNR mod 2 metallic nanoribbon. In the left panel
of Fig.~\ref{Optical Transitions aGNR}(c) we plot the absorption
spectrum for light polarized parallel to the nanoribbon length.  The
total absorption spectrum, shown as a solid black line, is a sum of
absorption spectra due to three major interband direct transitions
labeled $E_{11}$, $E_{22}$ and $E_{33}$.  In the metallic armchair
nanoribbons, we follow convention and label optical transitions
starting with the lowest one defined as $E_{11}$. For clarity,
transition energies are sometimes labeled $E^s_{ii}$ or $E^m_{ii}$ in
semiconducting and metallic nanoribbons, respectively. Our notation
should cause no confusion since it is clear from the context whether
we are talking about semiconducting or metallic nanoribbons.  From the
band diagram in Fig.~\ref{Optical Transitions aGNR}(c) we see that
$E_{11}$ corresponds to a direct transition between the second
electron and hole bands. Optical transitions between the first
electron and hole bands with linear dispersion don't contribute to the
absorption spectrum. They do, however, pin the Fermi level as a
function of T at $E_{\rm F} = 0 \ \mbox{meV}$ as can be seen in
Fig.~\ref{Optical Transitions aGNR}(c).

%------------------------------------------------------------------------------
\subsection{Phonon dispersion relations}
%------------------------------------------------------------------------------
%
%..............................................................................
% Phonon dispersion relations for aGNR nanoribbon
%..............................................................................
\begin{figure} [tbp]
\includegraphics[scale=.75]{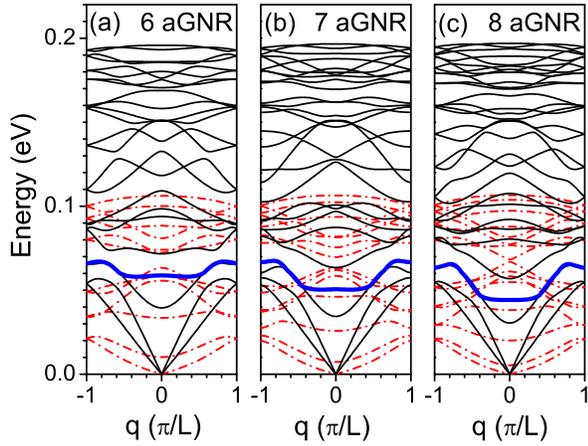}
\caption{ (color online) Phonon dispersion relations for aGNR
  nanoribbons calculated in the valence force field model for (a)
  6~$\beta$-aGNR mod 0, (b) 7~$\gamma$-aGNR mod 1, and (c)
  8~$\alpha$-aGNR mod 2 nanoribbons.  The out-of-plane modes are red
  dash-dotted lines while the remainder are in-plane modes. The
  dispersion band associated with the $q=0$ RBLM mode is shown as a
  thick blue line.  }
\label{aGNR Phonon Bands}
\end{figure}
%..............................................................................

Phonon dispersion relations for 6~$\beta$-aGNR mod 0, 7~$\gamma$-aGNR
mod 1, and 8~$\alpha$-aGNR mod 2 nanoribbons obtained in our VFF model
are shown in Figs.~\ref{aGNR Phonon Bands}(a), (b) and (c),
respectively. As in the zigzag nanoribbon case, dispersion relations
and normal mode displacements are obtained by solving the dynamical
matrix eigenvalue problem.  The in-plane and out-of-plane modes can
again be treated independently.  There are $2\ N_{ab}$ out-of-plane
modes (red dash-dotted lines) and $4\ N_{ab}$ in-plane modes (black
solid lines). In all cases, we find that $N_{ab}$ of the in-plane
$q=0$ modes can be coherently excited. The lowest of these modes is an
RBLM mode. The RBLM phonon dispersion relations are shown in
Fig.~\ref{aGNR Phonon Bands} as thick blue lines.

The $q=0$ RBLM phonon energies in aGNR nanoribbons obtained from our
VFF model can be fit to $\hbar\omega = A W^P + B$ where $W$ is the
armchir nanoribbon width in Angstroms and $\hbar\omega$ is the $q=0$
RBLM phonon energy in eV.  The fitting parameters $A = 0.31376$, $P =
-0.95579$ and $B = 0.00328$ are obtained from a least squares fit to
our VFF results for $4 \leq N_{ab} \leq 25$ corresponding to widths in
the range $4\unitangstrom < W < 30 \unitangstrom$.

%------------------------------------------------------------------------------
\subsection{Generation of coherent phonons}
%------------------------------------------------------------------------------

In our simulation, we excite coherent RBLM phonons with a 20~fs
Gaussian laser pulse. The fluence F is $5 \times 10^4
\ \mbox{J/cm}^2$ and the pump and probe pulse are linearly polarized with an
electric polarization vector parallel to the nanoribbon length. Time
dependent distribution functions are obtained by integrating the
photogeneration rates in Eq.~(\ref{Photogeneration rate}) using the
equilibrium distribution functions as initial conditions. We assume a
FWHM pump spectral width of 50~meV which appears in the broadened
delta function in Eq.~(\ref{Photogeneration rate}). Using the
time-dependent distribution functions, the RBLM coherent phonon
driving function $S(t)$ can be obtained from Eq.~(\ref{Coherent phonon
  driving function}).
%
%..............................................................................
% CP spectra for 6 aGNR nanoribbon
%..............................................................................
\begin{figure*} [tbp]
\includegraphics[width=18cm]{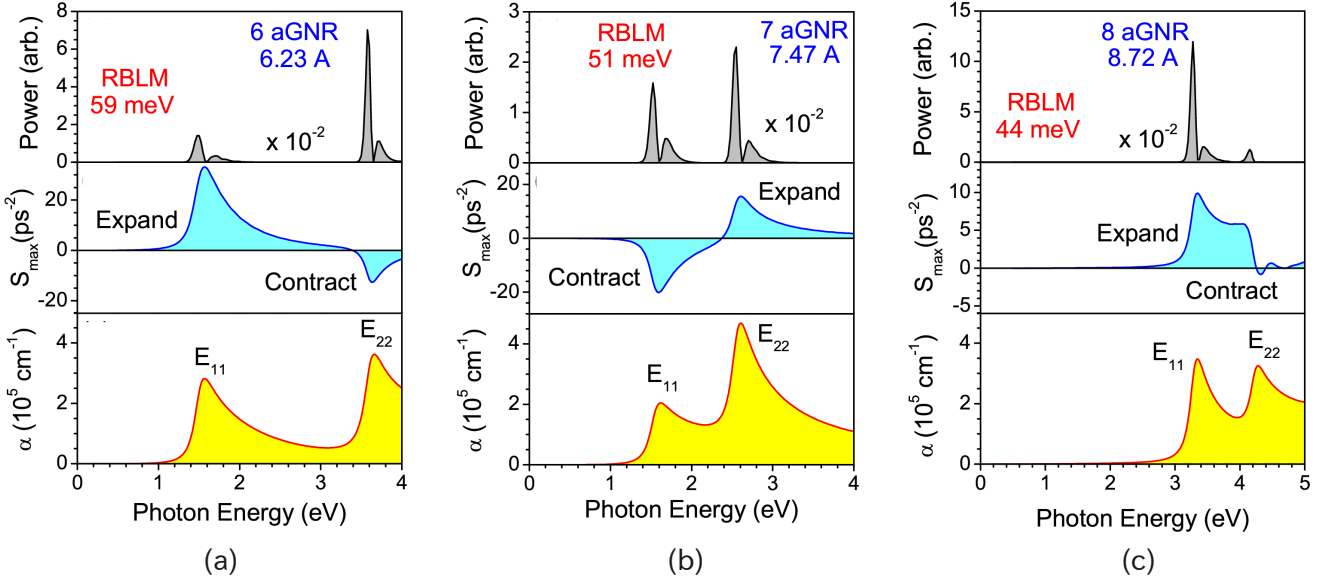}
\caption{ (color online) The coherent phonon power, the value of
  $\Smax$, and the initial absorption spectrum are plotted as a
  function of photon energy for (a) 6~$\beta$-aGNR mod 0 nanoribbon
  (RBLM frequency = 59~meV) , (b) 7~$\gamma$-aGNR mod 1 nanoribbon
  (RBLM frequency = 51~meV), and (c) 8~$\alpha$-aGNR mod 2 nanoribbon
  (RBLM frequency = 44~meV).  The excitation is due to a Gaussian
  laser pulse with pump and probe polarization vectors parallel to the
  ribbon length }
\label{aGNR CP_Power}
\end{figure*}
%..............................................................................

We examine $\Smax$ as a function of pump photon energy and our results
for 6~$\beta$-aGNR mod 0 semiconducting nanoribbons are shown in
Fig.~\ref{aGNR CP_Power}(a) where $\Smax$ is shown as a function of
pump photon energy. For comparison, the absorption coefficient is also
plotted in the lower panel of Fig.~\ref{aGNR CP_Power}(a).

Near the band edge, we see from Fig.~\ref{aGNR CP_Power}(a) that the
pump light is strongly absorbed near the $E_{11}$ and $E_{22}$
peaks. The resulting increase in the photoexcited carrier density
increases the coherent phonon driving function and enhances the
coherent phonon oscillation amplitudes.  Photoexcitation by the pump
causes the nanoribbon to initially expand for pump photon energies
near the $E_{11}$ transition and to initially contract for pump photon
energies near the $E_{22}$ transition. We find this to be true for all
$\beta$-aGNR mod 0 semiconducting nanoribbons.

Qualitatively different results are obtained for $\gamma$-aGNR mod 1
nanoribbons.  In Fig.~\ref{aGNR CP_Power}(b) we plot $\Smax$ as a
function of pump photon energy for a 7~$\gamma$-aGNR mod 1 nanoribbon
and find that photoexcitation by the pump causes the nanoribbon to
initially contract for photon energies near the $E_{11}$ peak and
initially expand for photon energies near the $E_{22}$ peak. This is
found to be true for all $\gamma$-aGNR mod 1 semiconducting
nanoribbons.

In Fig.~\ref{aGNR CP_Power}(c) we show results for an 8~$\alpha$-aGNR
mod 2 metallic nanoribbon excited by a laser pulse polarized parallel
to the ribbon length. From Fig.~\ref{aGNR CP_Power}(c), we see that
photoexcitation by the pump causes the nanoribbon to initially expand
for photon energies near the $E_{11}$ transition. For photon energies
near the $E_{22}$ transition, the situation is more ambiguous.
%
%..............................................................................
% CP intensity vs polarization for 7 aGNR nanoribbon
%..............................................................................
\begin{figure} [tbp]
\includegraphics[scale=.75]{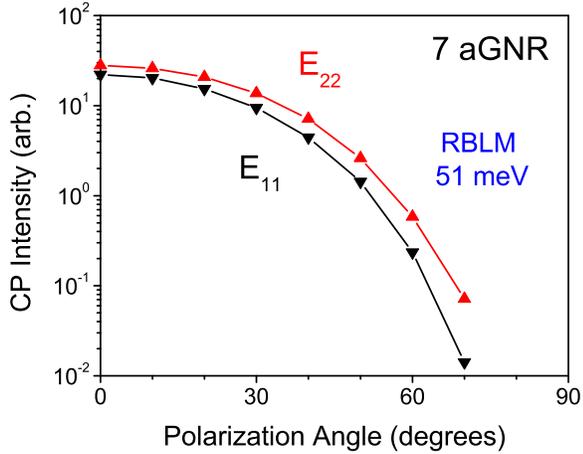}
\caption{ (color online) Integrated RBLM CP intensity for $E_{11}$ and
  $E_{22}$ transitions excited by a Gaussian laser pulse as a function
  of pump/probe polarization angle in a 7~$\gamma$-aGNR mod 1
  nanoribbon.  }
\label{a7GNR CP_Polarization}
\end{figure}
%..............................................................................

We have studied the CP intensity for ultrafast photoexcitation near
the $E_{11}$ and $E_{22}$ transitions in a 7~$\gamma$-aGNR mod 1
semiconducting nanoribbon as a function of polarization angle. In our
simulation, we assume the polarization of pump and probe are parallel
with the pump/probe polarization angle making an angle $\theta$ with
the nanoribbon axis. Our results are shown in Fig.~\ref{a7GNR
  CP_Polarization} and we can see that CP intensity for the $E_{11}$
and $E_{22}$ features in 7~$\gamma$-aGNR mod 1 semiconducting
nanoribbons drop off sharply as a function of polarization angle.
%
%..............................................................................
% CP intensity vs ribbon width for aGNR nanoribbons
%..............................................................................
\begin{figure} [tbp]
\includegraphics[scale=1.]{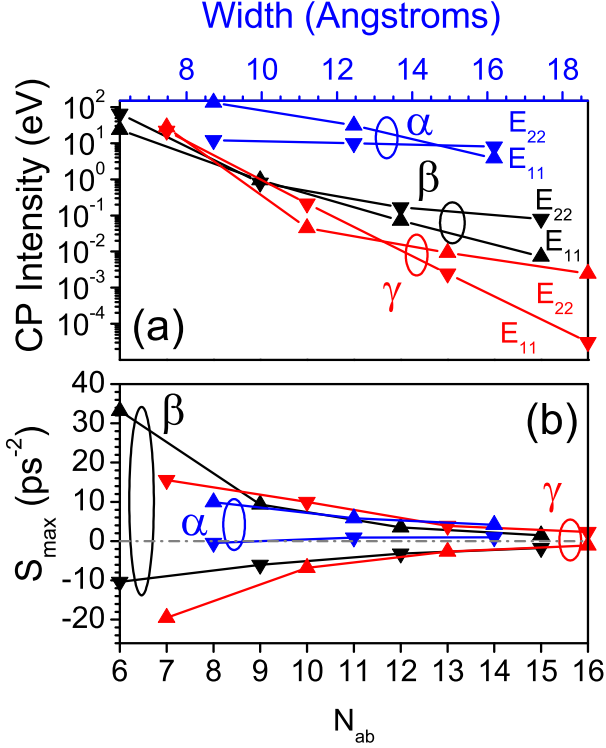}
\caption{ (color online) For $\alpha$-, $\beta$-, and $\gamma$-aGNR
  nanoribbons excited by Gaussian laser pulses with polarization
  vector parallel to ribbon length, we plot (a) RBLM CP intensity for
  $E_{11}$ and $E_{22}$ transitions and (b) the corresponding coherent
  phonon driving functions $S_{max}$ as functions of $N_{ab}$, the
  number of carbon dimers in the zigzag unit cell. The curves with
  upward pointing arrows are for the $E_{11}$ transitions and the
  downward pointing arrows are for the $E_{22}$ transitions. The ribbon width
  for each value of $N_{ab}$ can be read from the upper axis.}
\label{aGNR CP_Width dependence}
\end{figure}
%..............................................................................

We simulated CP intensity as a function of $N_{ab}$ in armchair
nanoribbons photoexcited by ultrafast Gaussian laser pulses polarized
parallel to the nanoribbon length. The results are shown in
Fig.~\ref{aGNR CP_Width dependence}(a) for the $E_{11}$ and $E_{22}$
features in $\alpha$-, $\beta$-, and $\gamma$-aGNR nanoribbons. The
corresponding values of the coherent phonon driving functions $S_{max}$
are shown in Fig.~\ref{aGNR CP_Width dependence}(b).

%%%%%%%%%%%%%%%%%%%%%%%%%%%%%%%%%%%%%%%%%%%%%%%%%%%%%%%%%%%%%%%%%%%%%%%%%%%%%%%
\section{Analysis of the electron-phonon interaction}
\label{EffMassTheory}
%%%%%%%%%%%%%%%%%%%%%%%%%%%%%%%%%%%%%%%%%%%%%%%%%%%%%%%%%%%%%%%%%%%%%%%%%%%%%%%

Following our previous study,~\cite{nugraha11-cpprb} in this section
we examine the $k$-dependent electron-phonon interaction in the
effective mass approximation to explain why some GNRs start their
coherent RBLM oscillations by initially expanding while others start
the oscillations by initially shrinking.  In the present discussion we
will focus our attention on the aGNRs, in which we can directly use
the wavefunctions formulated in previous
papers.~\cite{sasaki08-PTPs,nugraha11-cpprb} In the case of zGNRs, we
have to consider a special localized wavefunction due to the presence
of edge states at which the $E_{ii}$ transition
occurs.~\cite{sasaki06-edge} Using such a wavefunction, we obtain a
constant electron-phonon matrix element that does not depend on
$\mod(N_{ab},3)$ of the zGNRs, and thus is consistent with our results in
Sec.~\ref{Zigzag nanoribbon results section}.  The details will be
presented elsewhere.

The electron-phonon matrix element $\Melph$ is a sum of conduction
band (${\rm c}$) and valence band (${\rm v}$) electron-phonon matrix
elements, which represent the electron and hole contributions,
respectively,
\begin{align}\label{eq:elph}
\Melph &= \Melph^{\rm c} - \Melph^{\rm v} \notag \\
&= \langle {\rm c} | \Helph | {\rm c} \rangle
- \langle {\rm v} | \Helph | {\rm v} \rangle,
\end{align}
where $\Helph$ is the GNR electron-phonon interaction Hamiltonian and
the minus sign comes from the opposite charges of electrons and
holes.  In a nearest-neighbor effective mass approximation, the RBLM
$\Helph$ for an aGNR can be written as
\begin{equation}\label{eq:armelph}
\Helph = \uarm
\begin{pmatrix}
\gon & -\frac{\goff}{2} \\
-\frac{\goff}{2} & \gon
\end{pmatrix},
\end{equation}
where $\gon$ ($\goff$) is the on-site (off-site) coupling constant in
$\unitev$, while $\uarm$ is a ribbon width- or $N_{ab}$-dependent
phonon amplitude.  See Appendix~\ref{app:elph} for the derivation of
this Hamiltonian.  Next, to obtain $\Melph$ in Eq.~(\ref{eq:elph}), we
use the following wavefunctions,~\cite{sasaki08-PTPs}
\begin{equation}
\label{eq:wf}
\Psi_{\rm c} = \frac{\euler^{\imag\bf{k} \cdot \bf{r}}}{\sqrt{2S}}
\begin{pmatrix}
\euler^{-\imag\Theta(\bf{k})/2} \\
\euler^{+\imag\Theta(\bf{k})/2}
\end{pmatrix},
\Psi_{\rm v} = \frac{e^{i\bf{k} \cdot \bf{r}}}{\sqrt{2S}}
\begin{pmatrix}
\euler^{-\imag\Theta(\bf{k})/2} \\
-\euler^{+\imag\Theta(\bf{k})/2}
\end{pmatrix},
\end{equation}
for conduction and valence states, respectively.  In Eq.~(\ref{eq:wf}),
$S$ is the surface area of
graphene and $\thk$ is an angle of ${\bf k} = (k_x, k_y)$ measured
from the $k_x$-axis (ribbon width direction).
These wavefunctions are suitable near the graphene Dirac $K$ point and
thus they can explain well the aGNR lattice response especially at
relatively low energy $E_{11}$ and $E_{22}$ optical transitions.

By inserting the wavefunctions in Eq.~(\ref{eq:wf}) into
Eq.~(\ref{eq:elph}), we obtain
\begin{subequations}
\begin{align}
\langle {\rm c} | \Helph | {\rm c} \rangle &= \uarm \left(2 \gon
-\goff \cos \thk \right),\label{eq:effmass1stelec}\\
\langle {\rm v} | \Helph | {\rm v} \rangle &= \uarm \left(2 \gon
+\goff \cos \thk \right),\label{eq:effmass1sthole}
\end{align}
\end{subequations}
and thus
\begin{equation}
\Melph = \uarm \left( -2 \goff \cos \thk\right).
\label{eq:effmass1st}
\end{equation}
From this equation, we can analyze the $N_{ab}$ and $E_{ii}$
dependence of the aGNR initial lattice response.  First of all, we
should note that $\goff$ (which is usually taken to be $6.4~\unitev$
as mentioned in Ref.~\onlinecite{nugraha11-cpprb}) and $\uarm$
(cf. Eq.~(\ref{eq:uarm})) are always positive, while $\cos \thk$ can
either be positive or negative depending on the value of $\bf k$ at which the
$E_{ii}$ transition occurs.
%
%..............................................................................
% cutting line of mod 1 and mod 2 aGNR
%..............................................................................
\begin{figure} [tbp]
\includegraphics[clip,width=8cm]{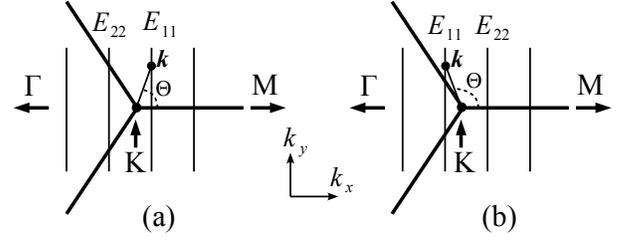}
\caption{\label{fig:mod0mod1aGNR} Cutting lines for (a)
  mod~0 $\beta$-aGNR and (b) mod~1 $\gamma$-aGNRs near the Dirac $K$ point.
  To make clear the definition of $\thk$, in this figure $\thk$ is shown for an
  arbitrary ${\bf k}$ at $E_{11}$.  In fact, in the case of mod~0 and
  mod~1 aGNRs the $E_{11}$ transitions occur at $\thk = 0$ and $\thk =
  \pi$, respectively.  The difference between the mod~0 and mod~1
  aGNRs can be understood from the position of the $E_{11}$ or
  $E_{22}$ cutting lines relative to the $K$ point. }
\end{figure}
%..............................................................................

\begin{figure} [tbp]
\includegraphics[scale=.75]{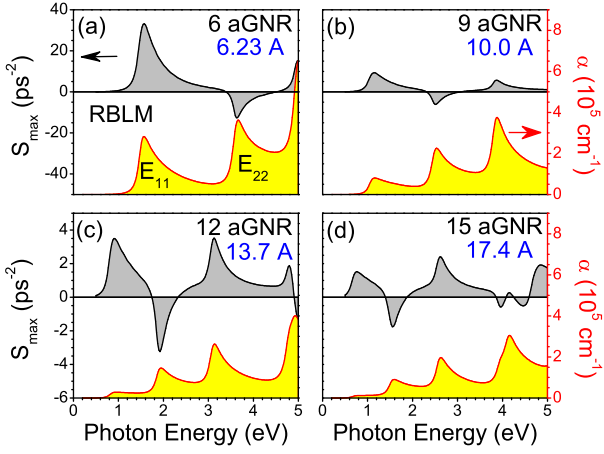}
\caption{\label{fig:aGNRSmaxmod0} (color online) Driving force $\Smax$
  and initial absorption spectrum as a function of photon energy for
  several mod~0 semiconducting $\beta$-aGNRs: (a) $N_{ab} = 6$, (b) $N_{ab} =
  9$, (c) $N_{ab} = 12$, and (d) $N_{ab} = 15$.  Positive (negative)
  $\Smax$ at $E_{11}$ ($E_{22}$) corresponds to an expansion
  (contraction) of the ribbon width.  $\delta E = E_{22} - E_{11}$
  is also shown, which is decreasing as a function of $N_{ab}$.}
\end{figure}

\begin{figure} [tbp]
\includegraphics[scale=.75]{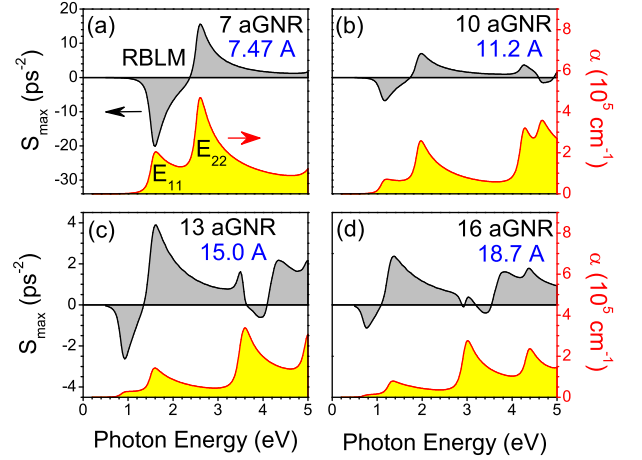}
\caption{\label{fig:aGNRSmaxmod1} (color online) Driving force $\Smax$
  and initial absorption spectrum as a function of photon energy for
  several mod~1 semiconducting $\gamma$-aGNRs: (a) $N_{ab} = 7$, (b) $N_{ab} =
  10$, (c) $N_{ab} = 13$, and (d) $N_{ab} = 16$.  Negative (positive)
  $\Smax$ at $E_{11}$ ($E_{22}$) corresponds to a contraction
  (expansion) of the ribbon width.}
\end{figure}

According to the definition of the driving force in
Eqs.~(\ref{Coherent phonon driven oscillator equation}) and
(\ref{Coherent phonon driving function}), a negative (positive)
$\Melph$ value corresponds to a positive (negative) $\Smax$.
Therefore, a positive (negative) $\cos \thk$ is related to a
contraction (expansion) of the ribbon width.  Using this argument, we
can classify the aGNR lattice response based on the aGNR types. For
example, let us consider semiconducting mod~0 $\beta$-aGNR and mod~1 $\gamma$-aGNRs.
The cutting line position for their $E_{11}$ and $E_{22}$ optical
transitions are just opposite to each other. For a mod~0 $\beta$-aGNR, we see
that $\cos \thk$ becomes positive (negative) at $E_{11}$ ($E_{22}$),
and thus the aGNR starts the coherent phonon oscillations by expanding
(shrinking) its width.  This can be seen in the illustration of $\thk$ in
Fig.~\ref{fig:mod0mod1aGNR}.  The opposite behavior is true for mod~1
$\gamma$-aGNRs.  These arguments are consistent with our numerical
results in Figs.~\ref{fig:aGNRSmaxmod0} and \ref{fig:aGNRSmaxmod1},
in which we show the aGNR driving force trends within the mod~0 $\beta$ family
and mod~1 $\gamma$ family, respectively.
%
%..............................................................................
% armchair driving force trends for mod 0 case
%..............................................................................
%..............................................................................
%
%..............................................................................
% armchair driving force trends for mod 1 case
%..............................................................................
%..............................................................................

\begin{figure} [tbp]
\includegraphics[scale=.75]{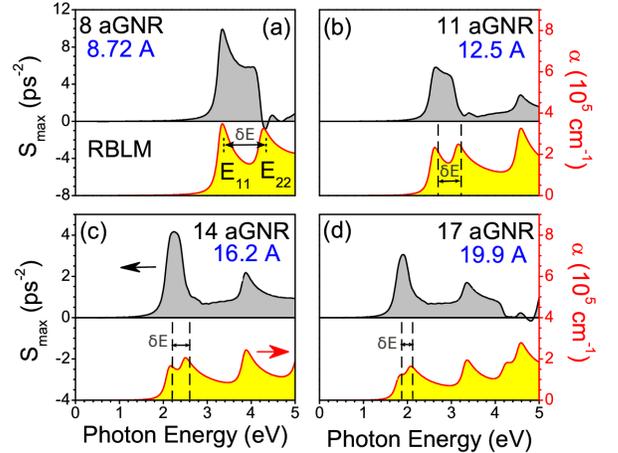}
\caption{\label{fig:aGNRSmaxmod2} (color online) Driving force $\Smax$
  and initial absorption spectrum as a function of photon energy for
  several mod~2 metallic aGNRs: (a) $N_{ab} = 8$, (b) $N_{ab} = 11$,
  (c) $N_{ab} = 14$, and (d) $N_{ab} = 17$.  Positive (negative)
  $\Smax$ at $E_{11}$ ($E_{22}$) corresponds to an expansion
  (contraction) of the ribbon width.}
\end{figure}

However, the driving force trends for the mod~2 metallic $\alpha$-aGNRs (see
Fig.~\ref{fig:aGNRSmaxmod2}) cannot be explained nicely by the
effective mass theory for several reasons.  The main reason is that,
in the metallic $\alpha$-aGNRs, there are two cutting lines with the same
distance from the $K$ point, which can be assigned as the lower and
higher branches of an $E_{ii}$ transition.  Both branches contribute
to a specific $E_{ii}$ and thus we have to sum up the matrix elements
from each contribution to obtain $\Melph$.  For example, if the 1D
$k$-points for the lower and higher branches of $E_{ii}$ are the same,
the matrix elements will cancel each other because $\cos\thk +
\cos(\pi - \thk) = 0$.  In this case, the CP amplitude will be generally small
for the mod~2 metallic aGNRs compared to the mod~0 or mod~1
semiconducting aGNRs.  In the real case, we always have slightly
different $k$-points for the lower and higher branches due to the
trigonal warping effect,~\cite{saito00-trig} from which the nonzero
$\Melph$ value gives information about an expansion or contraction of
the ribbon width.
%
%..............................................................................
% armchair driving force trends for mod 2 case
%..............................................................................
%..............................................................................

Near the $E_{11}$ transition, the metallic $\alpha$-aGNR initial
lattice response is always an expansion for all $N_{ab}$.  On the
other hand, near the $E_{22}$ transition, the response is expected to
be always a contraction, though we see in Figs.~\ref{fig:aGNRSmaxmod2}
(b)-(d) the trends is not true for larger $N_{ab}$.  We notice that
the difference $\delta_{E} = E_{22} - E_{11}$ might determine whether
or not the lattice response at the $E_{22}$ feature of a given mod~2
$N_{ab}$ $\alpha$-aGNRs will clearly follow our effective mass theory.
We guess that if $\delta_{E}$ is large enough ($\approx 2\unitev$ as in the
8~$\alpha$-aGNR), the lattice response at $E_{22}$ should not be
ambiguous. However, $\delta_{E}$ decreases with increasing
$N_{ab}$ as can be understood from a cutting line
argument~\cite{wakaba10-gnr}. Thus the lattice response at $E_{22}$ for
the larger mod~2 aGNRs becomes opposite to that for the smaller mod~2
aGNRs (e.g. $N_{ab} = 2, 5, 8$).

We should also note that the $E_{11}$ and $E_{22}$ values of the
metallic $\alpha$-aGNRs are close to the order of $E_{33}$ and $E_{44}$ values
for the semiconducting aGNRs.  This gives another reason why the
effective mass theory cannot explain the metallic $\alpha$-aGNR
trends. In this energy region, the nearest-neighbor effective mass theory
should be extended to include longer-range nearest-neighbor
interactions,~\cite{sasaki08-PTPs} which is beyond the scope of this
paper.  Nevertheless, the present discussion on the effective mass
theory has already given physical insight for explaining the initial
lattice response of mod~0 and mod~1 semiconducting aGNRs.  Even in the
case of mod~2 metallic $\alpha$-aGNRs we can see that the effective mass theory
is still able to explain the lattice behavior near the $E_{11}$
region.  Finally, we summarize the lattice behavior at $E_{11}$ and
$E_{22}$ transitions for all families of aGNRs in
Table~{\ref{table1}}.

\begin{table}
\caption{\label{table1} Initial lattice behavior due to coherent phonon
oscillations at $E_{11}$ and $E_{22}$ in aGNRs.}

\begin{tabular}{c|c|c}
\hline
family & $E_{11}$ & $E_{22}$ \\
\hline
mod~0 & expand & contract \\
mod~1 & contract & expand \\
mod~2 & expand & expand or contract\\
\hline
\end{tabular}

\end{table}

%%%%%%%%%%%%%%%%%%%%%%%%%%%%%%%%%%%%%%%%%%%%%%%%%%%%%%%%%%%%%%%%%%%%%%%%%%%%%%%
\section{Summary}
\label{Summary section}
%%%%%%%%%%%%%%%%%%%%%%%%%%%%%%%%%%%%%%%%%%%%%%%%%%%%%%%%%%%%%%%%%%%%%%%%%%%%%%%

We have developed a microscopic theory for the generation and
detection of coherent phonons in graphene nanoribbons using an extended
tight-binding model for the electronic states and a valence force
field model for the phonons.  We consider zigzag and armchair ribbons
denoted $N_{ab} \ \textrm{zGNR}$ and $N_{ab} \ \textrm{aGNR}$,
respectively, where $N_{ab}$ is the number of AB carbon dimers in the
translational unit cell.  In coherent phonon spectroscopy, ultrafast
laser pulses generate electrons and holes in the conduction and
valence bands of a graphene nanoribbon. If the pulse duration is less
than the phonon oscillation period, the photogenerated carriers couple
to the phonons through the deformation potential electron-phonon
interaction and the nanoribbon lattice undergoes coherent macroscopic
phonon oscillations.  The coherent phonon amplitudes satisfy a driven
oscillator equation (derived from the Heisenberg equation) with a
driving function that depends on the electron-phonon interaction
matrix elements and the photoexcited carrier distribution
functions. For large laser spot sizes, only $q=0$ phonon modes can be
coherently excited and these modes are said to be CP active. Coherent
phonons are detected using a delayed probe pulse to measure the time
dependent oscillations in the differential transmission. Taking the
Fourier transform of the differential transmission with respect to
probe delay time, we obtain the coherent phonon spectrum as a function
of phonon frequency with peaks in the CP spectrum corresponding to
excited coherent phonon modes.

For both zGNRs and aGNRs there are $N_{ab}$ CP active modes that
vibrate in the plane of the nanoribbon. In all cases, the CP active
mode with the lowest frequency is the RBLM mode. For 20~fs Gaussian
laser pulses we ignore carrier relaxation effects and integrate
Eq.~(\ref{Photogeneration rate}), the photogeneration rate
obtained from Fermi's golden rule, to obtain the photoexcited carrier
distribution functions. Using these carrier distribution functions
we obtain the driving function for RBLM coherent phonons as a function
of the pump photon energy and polarization angle.
In all cases we find that the driving function $\Smax$ immediately after
photoexcitation is greatest for light polarized parallel to the
nanoribbon axis. For photoexcitation near the optical absorption edge,
we find that the coherent phonon driving function for the RBLM mode
is much larger for zigzag nanoribbons where $\Delta n \neq 0$ transitions
involving localized edge states provide the dominant contribution to the
CP driving function. The sign of $\Smax$ is interesting since it gives
phase information that can be measured in CP spectroscopy. In zigzag
nanoribbons, the ribbons initially expand for pump photon energies
near the V2C1 and V3V1 absorption features.  In armchair nanoribbons
the phase behavior is different for the $\alpha$, $\beta$ and $\gamma$
families where $\Delta n = 0$ transitions between zone folded quantum
confined states give rise to the CP driving function.  In $\beta$-aGNR
mod 0 semiconducting aGNRs the ribbon initially expands for pumping
near the $E_{11}$ absorption peak and contracts for pumping near
$E_{22}$.  In $\gamma$-aGNR mod 1 semiconducting aGNRs the ribbon
initially contracts for pumping near the $E_{11}$ absorption peak and
expands for pumping near $E_{22}$.  In the case of the metallic
$\alpha$-aGNR mod 2 ribbons, the ribbon is seen to expand for pumping
near the $E_{11}$ feature while the behavior near the $E_{22}$ feature
is ambiguous.  Such lattice responses are also discussed by
considering the electron-phonon interaction within an effective mass
theory.

%%%%%%%%%%%%%%%%%%%%%%%%%%%%%%%%%%%%%%%%%%%%%%%%%%%%%%%%%%%%%%%%%%%%%%%%%%%%%%%
%                          Acknowledgements
%%%%%%%%%%%%%%%%%%%%%%%%%%%%%%%%%%%%%%%%%%%%%%%%%%%%%%%%%%%%%%%%%%%%%%%%%%%%%%%
\begin{acknowledgments}
This work was supported by the National Science Foundation under grant Nos.
DMR-1105437 and OISE-0968405, the Office of Naval Research under grant No.
ONR-00075094, R.S. acknowledges MEXT grant (Ministry of Education, Japan, No. 20241023.)
\end{acknowledgments}

%%%%%%%%%%%%%%%%%%%%%%%%%%%%%%%%%%%%%%%%%%%%%%%%%%%%%%%%%%%%%%%%%%%%%%%%%%%%%%%
\appendix
\section{Electronic states in extended tight binding model}
\label{app:electronic states}
%%%%%%%%%%%%%%%%%%%%%%%%%%%%%%%%%%%%%%%%%%%%%%%%%%%%%%%%%%%%%%%%%%%%%%%%%%%%%%%

In treating electronic states in zigzag and armchair nanoribbons within the
ETB model, we assume that all bond lengths and bond angles between A and B
carbon atoms are the same as that in planar graphene. In addition, we
assume the ribbon edges are unpassivated. Since the nanoribbons are assumed
to be perfectly planar, there is no mixing of $\sigma$ and $\pi$ orbitals
and the two types of bonds can be considered separately. The $\sigma$ bonds
account for the mechanical properties of the ribbon while the $\pi$
bonds determine the optical properties near the band edge and play
the dominant role in the generation of coherent phonons.  We treat
the valence $\pi$ and conduction $\pi^{*}$ bands using
the third-nearest-neighbor extended tight binding (ETB) model
developed by Porezag \textit{et al.} for carbon compounds.\cite{porezag95}
The Hamiltonian and overlap matrix elements between $\pi$ orbitals on
different sites are parameterized by analytic expressions that depend
on the C-C bond length. These parameterizations were calibrated
using DFT results for a local orbital basis set in the local density
approximation (LDA) for a wide range of carbon compounds.

A graphene nanoribbon only has translational symmetry along its
length with period $L$ so the Brillouin zone is one dimensional with
a wavevector $k$ satisfying $|k| \leq \pi/L$. The tight binding
wavefunction for a $\pi$ electron with wavevector $k$ is given by
\begin{equation}\label{tight binding wavefunction}
\psi_k (\textbf{r}) = \sum_{s=A}^{B} \ \sum_{j=1}^{N_{ab}}
C_{sj}(k) \sum_{l=-\infty}^{\infty} e^{i (k L) l} \
\phi_{\pi}(\textbf{r}-\textbf{R}_{{sj}}^{l})
\end{equation}
where $\phi_{\pi}(\textbf{r})$ are atomic $2p_z$ orbitals,
$\textbf{R}_{sj}^{l}$ is the position of carbon atom $sj$ in the l-th
translational unit cell.

The tight binding expansion coefficients $C_{sj}(k)$ in
Eq.~(\ref{tight binding wavefunction}) satisfy the generalized
eigenvalue equation
\begin{equation}\label{generalized eigenvalue equation}
\sum_{s'j'} H_{jj'}^{ss'}(k) \ C_{s'j'}(k) =
E(k) \sum_{s'j'} S_{jj'}^{ss'}(k) \ C_{s'j'}(k) .
\end{equation}
The Hamiltonian matrix elements are given by
\begin{equation}\label{Hamiltonian matrix elements}
H_{jj'}^{ss'}(k) = \sum_{l} e^{i (k L) l} \
H_{pp\pi}^{CC} \left( \left|
\textbf{R}_{s'j'}^{l} - \textbf{R}_{s'j'}^{0}
\right| \right),
\end{equation}
where $H_{pp\pi}^{CC}(d)$ is the Hamiltonian matrix element between
two carbon $2p_z$ $\pi$ orbitals as a function of the distance $d$
between them. A similar expression holds for $S_{jj'}^{ss'}(k)$ with
$H_{pp\pi}^{CC}(d)$ replaced by the overlap matrix element
$S_{pp\pi}^{CC}(d)$. The ETB matrix elements vanish at a finite
cutoff distance, and the sum over $l$ in Eq.~(\ref{Hamiltonian matrix
elements}) includes all contributions out to third nearest neighbor
shells. Analytic expressions for $H_{pp\pi}^{CC}$ and
$S_{pp\pi}^{CC}$ can be found in Ref.~\onlinecite{porezag95}.

Solving Eq.~(\ref{generalized eigenvalue equation}) we obtain the
electronic states $E_{n}(k)$ and wavefunctions where $n=1 \ldots 2
N_{ab}$ is the subband index in order of increasing energy. The first
$N_{ab}$ levels are the valence bands and the remainder are
conduction bands.

%%%%%%%%%%%%%%%%%%%%%%%%%%%%%%%%%%%%%%%%%%%%%%%%%%%%%%%%%%%%%%%%%%%%%%%%%%%%%%%
\section{Phonon modes in valence force field model}
\label{app:phonon modes}
%%%%%%%%%%%%%%%%%%%%%%%%%%%%%%%%%%%%%%%%%%%%%%%%%%%%%%%%%%%%%%%%%%%%%%%%%%%%%%%

We treat lattice dynamics in zigzag and armchair nanoribbons using a Born-von
Karman valence force field model.~\cite{saito98-phys,jishi82-graphite,jishi93-phonon}
We let $\textbf{U}_{sj}^{l}(t)$ be the displacement from equilibrium of a
carbon atom at $\textbf{R}_{sj}^{l}$. The equations of motion are given by
\begin{equation}\label{U(t) equations of motion}
\frac{d \textbf{U}_{sj}^{l}(t)}{dt}=\sum_{s'j'l'}
\textbf{K}_{sjl;s'j'l'}\left(
\textbf{U}_{s'j'}^{l'}(t)-\textbf{U}_{sj}^{l}(t)\right).
\end{equation}
The force constant tensor $\textbf{K}_{sjl;s'j'l'}$ depends on the
distance between atoms $sjl$ the $s'j'l'$. In order to describe bond
twisting interactions involving four atoms it is necessary to retain
up to fourth nearest neighbor shell force constant tensors. If $\eta$
labels nearest neighbor shells and $d_\eta$ is the distance between
them, the force constant tensor connecting $\eta$-th neighbor atoms
is
\begin{equation}\label{Force constant tensor}
\textbf{K}_{sjl;s'j'l'} =
\begin{pmatrix}
 \phi_r^\eta \tau_x^2 + \phi_{ti}^\eta \tau_y^2 &
 (\phi_{ti}^\eta-\phi^\eta_r) \tau_x \tau_y & 0 \\
 (\phi_{ti}^\eta-\phi_r^\eta) \tau_x \tau_y & \phi_r^\eta \tau_y^2 +
 \phi_{ti}^\eta \tau_x^2 & 0 \\
 0 & 0 & \phi_{to}^\eta \\
\end{pmatrix},
\end{equation}
where $\phi_r^\eta$ is the $\eta$-th neighbor force constant
parameter in the radial bond-stretching direction and
$\phi_{ti}^\eta$ and $\phi_{to}^\eta$ are $\eta$-th neighbor force
constants for the transverse in-plane and out-of-plane
directions. For neighbor shells $\eta=1 \ldots 4$, we use the force
constants tabulated in Table~I of Ref.~\onlinecite{jishi93-phonon}
and reproduced in Table 9.1 of Ref.~\onlinecite{saito98-phys}. The
vector $\tau$ is a vector of unit length from atom $sjl$ to $s'j'l'$.

Taking the Fourier transform of the atomic displacements, we obtain
\begin{equation}\label{U(t) normal mode expansion}
\textbf{U}_{sj}^{l}(t) = \frac{1}{\sqrt{N_\Omega}} \sum_\textbf{q}
\textbf{U}_{sj}(q) \ e^{-i \textbf{q} \cdot \textbf{R}_{sj}^{l}}
\end{equation}
where $\textbf{U}_{sj}(q)$ are the normal mode displacement vectors
and $\textbf{q}$ is parallel to the nanoribbon axis. The phonon
energies and mode displacement vectors are obtained by solving the
dynamical matrix eigenvalue problem
\begin{equation}\label{Dynamical matrix problem}
\sum_{s'j'} \textbf{D}_{sj;s'j'}(q) \ \textbf{U}_{s'j'}(q) =
M \omega(q)^2  \ \textbf{U}_{sj}(q) ,
\end{equation}
where $M$ is the carbon atom mass and $\omega(q)$ is the
eigenfrequency of the phonon mode. In terms of the force constant
tensor, the dynamical matrix is given by
\begin{eqnarray}\label{Dynamical matrix} \nonumber
\textbf{D}_{sj;s'j'}(q) &=& \delta_{ss'} \delta_{jj'}
\sum_{s''j''l''} \textbf{K}_{sj0;s''j''l''} \\ &-& \sum_{l''}
\textbf{K}_{sj0;s'j'l''} \ e^{-i \textbf{q} \cdot \left(
\textbf{R}_{s'j'}^{l''}-\textbf{R}_{sj}^{0} \right) }.
\end{eqnarray}

For each value of $q$ the dynamical matrix eigenvalue problem can be
solved to obtain phonon energies $\hbar \omega_{m}(q)$ and
displacement vectors $\textbf{U}_{sj}^m(q)$ where $m$ labels the
phonon modes.

%%%%%%%%%%%%%%%%%%%%%%%%%%%%%%%%%%%%%%%%%%%%%%%%%%%%%%%%%%%%%%%%%%%%%%%%%%%%%%%
\section{Electron-phonon interaction in ETB model}
\label{app:elph ETB}
%%%%%%%%%%%%%%%%%%%%%%%%%%%%%%%%%%%%%%%%%%%%%%%%%%%%%%%%%%%%%%%%%%%%%%%%%%%%%%%

The second-quantized electron-phonon interaction in zigzag and armchair
nanoribbons in the ETB model is obtained by evaluating the integral
\begin{equation}\label{Electron phonon second quantization integral}
\hat{H}_{ep} = \int d\textbf{r} \
\hat{\psi}^{\dag}(\textbf{r}) \ H_{ep}(\textbf{r})
\ \hat{\psi}(\textbf{r})
\end{equation}
where the deformation potential electron-phonon interaction Hamiltonian is
\begin{equation}\label{Electron phonon Hamiltonian}
H_{ep}(\textbf{r}) = - \ \sum_{sjl}
\nabla v_c(\textbf{r}-\textbf{R}_{sj}^l) \cdot \textbf{U}_{sj}^{l}.
\end{equation}

The electron field operator $\hat{\psi}(\textbf{r})$ is
\begin{equation}\label{Electron field operator}
\hat{\psi}(\textbf{r}) = \sum_{nk} c_{nk}^{} \ \psi_{nk}(\textbf{r}),
\end{equation}
where $\psi_{nk}(\textbf{r})$ are the tight binding wavefunctions for
wavevector $k$ in Eq.~(\ref{tight binding wavefunction}) with $n=1
\ldots 2 N_{ab}$ labeling the subband levels. The annihilation
operator for this electronic state is $c_{nk}^{}$.

After carrying out this integral over the electronic coordinate, we
obtain the final result by replacing the classical atomic
displacements $\textbf{U}_{sj}^{l}$ with the second-quantized operator
\begin{equation}\label{U second quantized operator}
\textbf{U}_{sj}^{l} = \frac{\hbar}{\sqrt{2 \rho L_\Omega}} \sum_{mq}
\frac{\hat{\textbf{e}}_{sj}^m(q)}{\sqrt{\hbar\omega_m(q)}}
e^{i q L l} \left( b_{mq}^{}+b_{m,-q}^{\dag} \right).
\end{equation}
Here $\rho$ is the mass density per unit length and $L_\Omega$ is
the nanoribbon length. The atomic displacements $\textbf{e}_{sj}^m(q)$ are
eigenvectors of the dynamical matrix in Eq.~(\ref{Dynamical matrix}). They
satisfy the normalization condition
\begin{equation}\label{Displacment vector normalization condition}
\sum_{sj} \textbf{e}_{sj}^m(q)^{*} \cdot \textbf{e}_{sj}^{m'}(q)^{} =
\delta_{m,m'}.
\end{equation}

The final result for the second-quantized electron-phonon interaction is
\begin{equation}\label{Second quantized electron-phonon hamiltonian}
\hat{H}_{ep}=\sum_m \sum_{n'k';nk} M^{mq}_{n'k';nk} \
c_{n'k'}^{\dag} c_{nk}^{} \ (b_{mq}+b_{m,-q}^{\dag}),
\end{equation}
where $q \equiv k'-k$. The interaction matrix elements is given by
\begin{eqnarray}\label{Electron-phonon matrix element}
\nonumber &&
M^{mq}_{n'k';nk} = -A_m(q) \sum_{sj;s'j'}
C_{s'j'}^{*}(n'k') \ C_{sj}^{}(nk) \times
\\ &&
\sum_{ll';s''j''} e^{iL (kl-k'l')} \
\hat{\textbf{e}}_{s''j''}^m(q) \cdot \vec{\lambda}(s'j'l';s''j'';sjl),
\end{eqnarray}
where $A_m(q)=\hbar/\sqrt{2 \rho L_\Omega \ \hbar\omega_m(q)}$ is the
quantized phonon amplitude. In Eq.~(\ref{Electron-phonon matrix
element}) L is the length of the translational unit cell,
$C_{sj}^{}(nk)$ are the electron tight-binding expansion
coefficients in Eq.~(\ref{tight binding wavefunction}),
$\hat{\textbf{e}}_{sj}^m(q)$ are the atomic normal mode displacements
obtained by diagonalizing the dynamical matrix in Eq.~(\ref{Dynamical
matrix}), and $\vec{\lambda}$ is the deformation potential vector.

The deformation potential vector is the three center integral
\begin{equation}\label{Deformation potential vector}
\vec{\lambda} = \int d\textbf{r} \
\phi_{\pi}^{*}(\textbf{r}-\textbf{R}_{s'j'}^{l'}) \
\nabla v(\textbf{r} - \textbf{R}_{s''j''}^0) \
\phi_{\pi}^{}(\textbf{r}-\textbf{R}_{sj}^{l}).
\end{equation}
We evaluate $\vec{\lambda}$ using the $2p_z$ atomic wavefunctions
and screened atomic potential for carbon in
Ref.~\onlinecite{jiang05-elphint} obtained from an {\it ab initio}
calculation in graphene.~\cite{gruneis04-phd} The $\pi$ orbitals are
expanded as
\begin{equation}\label{Pi orbital}
\phi_\pi(\textbf{r}) =
z \ \sum_{l=1}^4 I_l \ \exp \left( -\frac{r^2}{2 \sigma_l^2} \right),
\end{equation}
where the $z$ direction is perpendicular to the nanoribbon plane.
Similarly the screened atomic potentials are expanded as
\begin{equation}\label{Atomic potential}
v(\textbf{r}) =
\frac{1}{r} \sum_{l=1}^4 v_l \ \exp \left( -\frac{r^2}{2 \tau_l^2}
\right).
\end{equation}
Values of $I_l$, $v_l$, $\sigma_l$ and $\tau_l$ are tabulated in Table
I of Ref.~\onlinecite{jiang05-elphint}. Substituting the expansions
(\ref{Pi orbital}) and (\ref{Atomic potential}) into
Eq.~(\ref{Deformation potential vector}) we obtain an expansion for
$\vec{\lambda}$ in terms of three-dimensional integrals which can be
evaluated analytically.\cite{jiang05-elphint,gruneis04-phd}

%%%%%%%%%%%%%%%%%%%%%%%%%%%%%%%%%%%%%%%%%%%%%%%%%%%%%%%%%%%%%%%%%%%%%%%%%%%%%%%
\section{RBLM electron-phonon interaction in effective mass model}
\label{app:elph}
%%%%%%%%%%%%%%%%%%%%%%%%%%%%%%%%%%%%%%%%%%%%%%%%%%%%%%%%%%%%%%%%%%%%%%%%%%%%%%%

\begin{figure}[htbp]
 \begin{center}
 \includegraphics[clip,width=8cm]{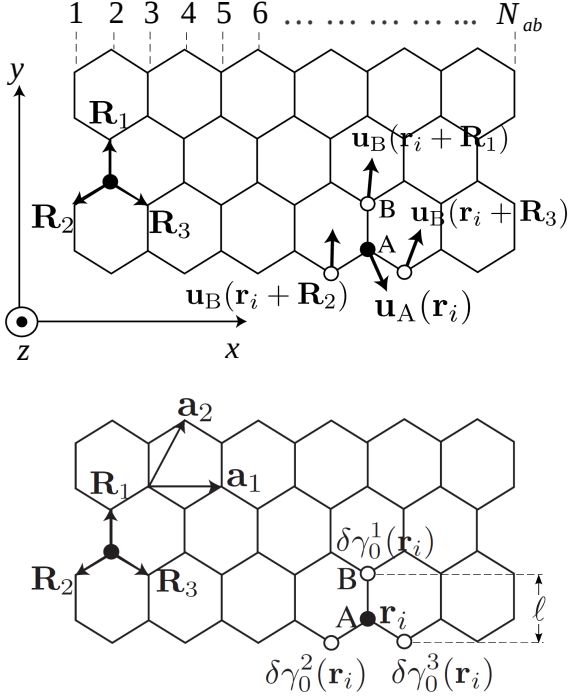}
 \end{center}
 \caption{\label{fig:graphene-rblm} Upper panel shows displacements of
   B-atoms at ${\bf r}_i + {\bf R}_a$ ($a=1,2,3$), that is ${\bf
     u}_{\rm B}({\bf r}_i + {\bf R}_a)$, which give rise to a
   deformation potential at A-atom of ${\bf r}_i$.  Lower panel shows
   local modulations of the hopping integral defined by
   $\dgamma0^a({\bf r}) $ ($a = 1,2,3)$.  In this coordinate system we
   have the nearest-neighbor vectors ${\bf R}_{1}= (0,\acc)$, ${\bf
     R}_{2}= (-\sqrt{3}/2,-1/2)\acc$, ${\bf R}_{3}=
   (\sqrt{3}/2,-1/2)\acc$, where $\acc = a/\sqrt{3}$.  Here $\ell =
   3\acc/2$ is used in Eq.~(\ref{eq:deltagamma}).}
\end{figure}

In this section, we derive Eq.~(\ref{eq:armelph}) which gives the RBLM
electron-phonon interaction Hamiltonian $\Helph$ for an armchair
nanoribbon.  The Hamiltonian can be decomposed into the on-site and
off-site Hamiltonians.  The details of the on-site and off-site
interactions are given in Sasaki's work on the deformation-induced
gauge field in graphene.~\cite{sasaki08-PTPs} We will directly use his
results in formulating the on-site and off-site Hamiltonians.  The
on-site and off-site interactions are induced by a lattice deformation
which gives rise to a change in the transfer integral and a change in
the potential between A and B atoms.  We adopt a coordinate system
shown in Fig.~\ref{fig:graphene-rblm} to derive $\Helph$.
%
%...........................................................................
% delta gamma figure
%...........................................................................
%..........................................................................

\subsection*{On-site Hamiltonian}
The on-site Hamiltonian can be expressed in terms of the divergence of
$\uA$ and $\uB$, which represent the displacement vector of A-atom
and B-atom in the graphene unit cell, respectively.  This Hamiltonian
is written as
\begin{equation}
 {\cal H}_{\rm on} = \gon
 \begin{pmatrix}
  \nabla \cdot \uB({\bf r}) & 0 \cr
  0 & \nabla \cdot \uA({\bf r})
 \end{pmatrix}.
 \label{eq:Hon_g}
\end{equation}
For the discussion of the RBLM electron-phonon interaction, we rewrite
Eq.~(\ref{eq:Hon_g}) as follows:
\begin{align}
{\cal H}_{\rm on} &= \gon \sigma_0 \nabla \cdot \left(\frac{\uA({\bf
    r}) + \uB({\bf r})}{2}\right) \notag \\ &+\gon \sigma_z \nabla
\cdot \left(\frac{\uA({\bf r}) - \uB({\bf r})}{2}\right),
 \label{eq:Honseparate}
\end{align}
where $\gon$ denotes the gradient of the atomic potential at ${\bf r}$,
$\sigma_0$ is the identity matrix, and $\sigma_z$ is the $z$-component of
the vector of Pauli matrices. For the RBLM in armchair nanoribbons
we have $\uA({\bf r}) = \uB({\bf r}) = {\bf u}({\bf r})$. Therefore,
Eq.~(\ref{eq:Honseparate}) can be simplified to be
\begin{equation}
{\cal H}_{\rm on} = \gon \sigma_0 \nabla \cdot \ur.
 \label{eq:Honsimple}
\end{equation}
The RBLM oscillation can be expressed by
\begin{equation}
\ur = u_m \sin (k_m x) \xhat,
 \label{eq:rblm}
\end{equation}
where $u_m$ is the maximum amplitude of the RBLM oscillation at the
armchair edge.  The wavevector $k_m$ is then specified by $\pi / W$,
where $W$ is the ribbon width.  By substituting $\ur$ in
Eq.~(\ref{eq:rblm}) into Eq.~(\ref{eq:Honsimple}), we obtain
\begin{equation}
{\cal H}_{\rm on} = \uarm
 \begin{pmatrix}
   \gon & 0 \\
   0 & \gon
 \end{pmatrix},
\end{equation}
where, for simplicity, here we have defined
\begin{equation}
\uarm (x) = k_m u_m \cos(k_m x) .
\label{eq:uarm}
\end{equation}
Note that $\uarm$ is always positive because the atomic position $x$
is within the range $-W/2 \leq x \leq W/2$.

\subsection*{Off-site Hamiltonian}
To derive the off-site interaction Hamiltonian, we start with the fact
that the lattice deformation modifies the nearest-neighbor hopping
integral locally as $-\gamma_0 \rightarrow -\gamma_0 + \dgamma0^a({\bf
  r}_i)$ ($a = 1,2,3$).  The corresponding perturbation of the lattice
deformation is given by
\begin{equation}
 {\cal H}_{\rm deform} \equiv
 \sum_{i \in {\rm A}} \sum_{a=1,2,3}
 \delta \gamma^a_0({\bf r}_i)
 \left[
 (c_{i+a}^{\rm B})^\dagger c_i^{\rm A} +
 (c_i^{\rm A})^\dagger c_{i+a}^{\rm B}
 \right],
 \label{eq:Hdeform}
\end{equation}
where $c_i^{\rm A}$ is the annihilation operator for a $\pi$ electron
on an A-atom at position ${\bf r}_i$, and $(c^{\rm B}_{i+a})^\dagger$
is the creation operator for a $\pi$ electron on a B-atom at
position ${\bf r}_{i+a}$ $(={\bf r}_i + {\bf R}_a)$.
This perturbation gives rise to scattering
within a region near the $K$-point of graphene whose interaction is
given by a deformation-induced gauge field ${\bf A}({\bf r})=(A_x({\bf
  r}),A_y({\bf r}))$ as $\vfermi \bsigma \cdot \left[ \hat{{\bf p}} +
  {\bf A}({\bf r}) \right]$, where $\vfermi$ is the Fermi velocity,
$\hat{{\bf p}} = -\imag\hbar\nabla$ is the momentum operator, and
$\bsigma$ is the Pauli matrix.  The deformation-induced gauge field
${\bf A}({\bf r})$ is defined from $\delta \gamma^a_0({\bf r})$
($a=1,2,3$) as~\cite{sasaki08-PTPs}
\begin{align}
 \begin{split}
  & v_{\rm F} A_x({\bf r}) = \delta \gamma^1_0({\bf r})
  - \frac{1}{2} \left[ \delta \gamma^2_0({\bf r}) +
  \delta \gamma^3_0({\bf r}) \right], \\
  & v_{\rm F} A_y({\bf r}) = \frac{\sqrt{3}}{2}
  \left[ \delta \gamma^2_0({\bf r}) -
  \delta \gamma^3_0({\bf r}) \right].
 \end{split}
 \label{eq:A}
\end{align}

Similar to the RBM electron-phonon interaction in carbon
nanotubes,~\cite{sasaki08-rbm} the perturbation to the
nearest-neighbor hoppping integral for the RBLM electron phonon
interaction in the armchair nanoribbons is given by
\begin{equation}
 \dgamma0^a({\bf r})
 = \frac{g_{\rm off}}{\ell \acc}
 {\bf R}_a \cdot
 \{ {\bf u}({\bf r}+{\bf R}_a)- \ur \},
 \label{eq:deltagamma}
\end{equation}
where $\goff$ is the off-site coupling constant and $\ell = 3\acc/2$ (see
the lower panel of Fig.~\ref{fig:graphene-rblm}).  Here the
displacement vector of a carbon atom at ${\bf r}$ in general is
expressed by $\ur = [u_x({\bf r}), u_y({\bf r})]$.  Using a Taylor
expansion, we approximate Eq.~(\ref{eq:deltagamma}) as
\begin{equation}
 \dgamma0^a({\bf r})
 = \frac{g_{\rm off}}{\ell \acc}
 {\bf R}_a \cdot
\left\{ ({\bf R}_a \cdot \nabla) \ur \right\}.
 \label{eq:deltagammaappr}
\end{equation}

Using ${\bf R}_1$, ${\bf R}_2$, and ${\bf R}_3$ in
Fig.~\ref{fig:graphene-rblm}, we obtain the deformation-induced gauge
field of Eq.~(\ref{eq:A}) as follows:
\begin{align}
 \begin{split}
  v_{\rm F} A_x({\bf r}) &= \frac{\goff}{2} \left[-\frac{\partial u_x
      ({\bf r})}{\partial x} + \frac{\partial u_y ({\bf r})}{\partial
      y} \right] , \\ v_{\rm F} A_y({\bf r}) &= \frac{\goff}{2}
  \left[\frac{\partial u_x ({\bf r})}{\partial y} + \frac{\partial u_y
      ({\bf r})}{\partial x} \right] .
 \end{split}
 \label{eq:Afinal}
\end{align}

Inserting ${\bf u}({\bf r})$ in Eq.~(\ref{eq:rblm}) into this last equation,
we then obtain
\begin{equation}
v_{\rm F} A_x = - \frac{\goff}{2} k_m u_m \cos(k_m x) = -
\frac{\goff}{2} \uarm,
\end{equation}
while $v_{\rm F} A_y = 0$.  Therefore, the off-site Hamiltonian can be
written as
\begin{equation}
{\cal H}_{\rm off} = \sigma_x v_F A_x = \uarm
 \begin{pmatrix}
   0 &  - \frac{\goff}{2} \\
    - \frac{\goff}{2} & 0
 \end{pmatrix}.
\end{equation}

Finally, we can get the electron-phonon interaction Hamiltonian of
Eq.~(\ref{eq:armelph}):
\begin{align}\label{}
\Helph &= {\cal H}_{\rm on} + {\cal H}_{\rm off} \notag\\
&= \uarm
\begin{pmatrix}
\gon & -\frac{\goff}{2} \\
-\frac{\goff}{2} & \gon
\end{pmatrix}.
\end{align}

\bibliography{paper}

\end{document}